\documentclass[aps,pra,floatfix,amsmath,amssymb,superscriptaddress,showpacs,showkeys,twocolumn,10pt]{revtex4-2}
\usepackage[caption=false]{subfig}
\usepackage{graphicx,bm,color}
\usepackage{amsfonts}
\usepackage{lipsum}
\usepackage{algorithm2e}
\usepackage[noend]{algpseudocode}
\makeatletter
\def\BState{\State\hskip-\ALG@thistlm}
\makeatother
\usepackage{color}
\usepackage{braket}
\usepackage[dvipsnames]{xcolor}
\usepackage{hyperref}
\hypersetup{
    colorlinks=true,
    linkcolor=blue,
    filecolor=blue,      
    urlcolor=blue,
}
\usepackage{epstopdf}

\begin{document}

\title{Characterization of Noise using variants of Unitarity Randomized Benchmarking}%

\author{Adarsh Chandrashekar}
\email{adarsh1chand@gmail.com}
\affiliation{Cryptology and Security Research Unit, Indian Statistical Institute, Kolkata, India}%
\affiliation{Currently working at Nykaa, Bengaluru, India}
\author{Soumya Das}
\email{soumya06.das@gmail.com}
\affiliation{Cryptology and Security Research Unit, Indian Statistical Institute, Kolkata, India}%
\affiliation{Currently working at Eindhoven University of Technology, Netherlands}
\author{Goutam Paul}
\email{goutam.paul@isical.ac.in}
\affiliation{Cryptology and Security Research Unit, Indian Statistical Institute, Kolkata, India}%

\begin{abstract}
Benchmarking noise induced during the implementation of quantum gates is the main concern for practical quantum computers. Several protocols have been proposed that empirically calculate various metrics that quantify the error rates of the quantum gates chosen from a preferred gate set. Unitarity randomized benchmarking (URB) protocol is a method to estimate the coherence of noise induced by the quantum gates which is measured by the metric \textit{unitarity}.  In this paper, we for the first time, implement the URB protocol in a quantum simulator with all the parameters and noise models used from a real quantum device. The direct implementation of the URB protocol in a quantum device is not possible using current technologies, as it requires the preparation of mixed states. To overcome this challenge, we propose a modification of the URB protocol, namely the m-URB protocol, that enables us to practically implement it on any quantum device. We validate our m-URB protocol using two single-qubit noise channels --  (a) depolarising channel and (b) bit-flip channel. We further alter the m-URB protocol, namely, native gate URB or Ng-URB protocol, to study the noise in the native gates into which the quantum circuits are compiled in a quantum computer. Using our Ng-URB protocol, we can also detect the presence of cross-talk errors which are correlated errors caused due to non-local and entangling gates such as CNOT gate.  For illustration, we simulate the noise of the native gates taking the noise parameter from two real IBM-Q processors which are of 5-qubit and 15-qubit quantum registers. Results show that there is a significant cross-talk in the CNOT gate, whereas the cross-talk is mild for the single-qubit gates, in both the processors. Our results highlight the practical effectiveness of the Ng-URB protocol in successfully detecting the cross-talk between qubits when applying native gate operations irrespective of the platform.
\end{abstract} 
\maketitle


\section{introduction}
\label{Introduction}
In the NISQ (Noisy-Intermediate Scale Quantum) era~\cite{NISQ}, researchers are primarily focused on what problems can be solved using a noisy quantum computer with the number of qubits in the order of $50-100$, which already surpass the abilities of a classical computer. Another direction of research in this era is towards achieving fault-tolerant quantum computation in the near future, i.e. to perform reliable computation either by using error-correction schemes or by studying and rectifying the sources of error in a quantum processor. In error-correction schemes, error-free computation is achieved by performing computation on `logical' qubits which are individually made of a number of physical noisy qubits present in a quantum processor. However, to perform practically significant computation the number of physical qubits required to create a single `logical' qubit may be large. An estimate in~\cite{Campbell} suggests that the number can be as high as $1000-10,000$ physical qubits for achieving fault tolerance in intermediate-scale processors. On the other hand, considerable research has been done into studying the noise induced in engineered quantum systems such as a quantum computer. 

When studying noisy quantum circuits, one can identify different forms of error introduced at various stages of quantum computation such as SPAM (state preparation and measurement) errors and gate errors~\cite{Decoherence,NotAllQubitsAreEqual}. To quantify the gate errors, several techniques have been proposed which are mostly clubbed under \textit{randomized benchmarking protocols}. Other methods have also been proposed such as quantum volume~\cite{QV} which indicates how faithfully a random quantum circuit of a given width and depth can be implemented in a quantum computer. Recently another metric has been proposed to benchmark the overall `quality' of quantum computation performed on actual devices via the generative capabilities of shallow quantum circuits executed on it~\cite{generative}. Finally, IBM has recently defined three key attributes for quantum computing performance: quality, speed, and scale which are relevant in the NISQ era. In order to quantify the `quality' of quantum computing performance, the class of randomized benchmarking protocols represents simple to-implement statistical methods that can be more informative about the different sources of error as compared to other holistic metrics such as quantum volume.
\subsection{Related Works}
The first of the randomized benchmarking protocols introduced was the \textit{Standard Randomized Benchmarking} (SRB)~\cite{Emerson, Knill,Helsen22,Proctor22,Nikolov23,Morvan2021,Xu2022,mehrani2024restricted,arienzo2024bosonic,PRXQuantum.5.030334}.  In this protocol, the metric used to quantify the gate error is the \textit{infidelity} of the noise channel. Infidelity is calculated from the fidelity measurement with respect to the identity channel of a quantum channel arising from the average noise associated with a given gate set (preferably a unitary 2-design~\cite{Dankert}). The quantum channel is constructed by applying a sequence of noisy gates from the gate set with the final gate inverting all the random gates applied before it, so that ideally (i.e., without noise) the quantum channel should be ab identity channel. Several variants of the SRB protocol have been proposed that benchmark the same metric but in different contexts~\cite{Magesan, Hashagen, Harper, Brown, Proctor}. 

\textit{Interleaved randomized benchmarking} (IRB)~\cite{Magesan} protocol benchmarks the average error rate for a given computational gate as compared to the average error for a gate set. Other variants of IRB can be found in~\cite{RB-individual-gates, efficient-RB}.

\textit{Real randomized benchmarking} (RRB) measures the average gate infidelity for a restricted gate set that is not a unitary 2-design~\cite{Hashagen, Harper, Brown}.

\textit{Direct randomized benchmarking} (DRB) benchmarks the performance of a gate set made of native gates present in a given quantum computing platform, i.e., the gates which have specific default instructions for their physical implementation readily available in the platform and which together form a gate set capable to perform universal quantum computation. This gate set usually does not form a unitary 2-design (or its subgroup)~\cite{Proctor}, etc.~\cite{Dirkse, Wallman,2302.13853}.

In general SRB protocols are based on the assumption of gate-independent noise. However, methods for analyzing the noise of a gate set without this assumption is studied in~\cite{gate-dependent, what-it-actually-measures}. SRB protocol for benchmarking noise in logical qubits, i.e., quantum processors with error correction schemes at the lowest level of its architecture has also been proposed in~\cite{logical-benchmarking}.  However, none of the actual quantum processors employ error correction principles for reliable quantum computing.

In SRB or its variants, infidelity quantifies the average error rate induced by the noisy gates. However, infidelity is unable to characterize the noise in terms of its \textit{coherence}, i.e., the extent to which the noise acts as a unitary channel. Unitarity~\cite{Wallman} is a metric to measure the coherence in an arbitrary noise channel. \textit{Unitarity randomized benchmarking} (URB)~\cite{Dirkse, Wallman,Girling22,Jafarzadeh20,Mitsuhashi24} protocol benchmarks the quality of a noise channel based on this metric.

For a unitary noise channel, the upper bound on the worst-case error rate with respect to diamond norm scales as the square root of the average gate infidelity, whereas for Pauli channels (which are non-unitary channels) it scales linearly~\cite{Sanders}. Since, based on the coherence of the noise, the actual error per gate scales with average infidelity differently for different noise channels, to that extent the effectiveness of the average infidelity to quantify the `quality' of gates applied in a quantum circuit is also limited. In particular, one can only qualitatively know whether the noise channel is unitary or not based on the scaling of the worst case error rate with the average infidelity, but the protocol does not provide a metric to `quantify' the unitary nature of a noise channel. 

Unitarity is one such metric calculated via the URB protocol. Such a metric is useful in benchmarking the `quality' of various physical implementations of a specific gate in a given processor (or across processors). It can also be useful for example, in quantitatively studying the effect of the connectivity between qubits in a processor, on the incoherence introduced in a multi-qubit gate by calculating the unitarity of the gate applied to different qubit subsystems. For these reasons, we are interested in the characterization of noise based on the coherence of the errors induced by different noise channels and therefore we hope that URB can aid in investigating the gate errors in practical quantum computation.

Another interpretation of unitarity is that it is the variance of the gate fidelity with respect to a gate set drawn from a unitary 2-design~\cite{interpretation-unitarity}. However, in this paper, we follow the definition by Wallman \textit{et al.}~\cite{Wallman}, since we later propose two modifications to the URB protocol which does not require a gate set but still follows the general definition of unitarity as proposed in~\cite{Wallman}. 
\subsection{Significant randomized benchmarking Experiments Performed in IBMQ}
Several quantum computing platforms with different archetypes, i.e., the way qubits are represented and manipulated are available today~\cite{Sycamore, Kjaergaard20, IonQ, TrappedIon, IBMQ}. For example, Google's Sycamore~\cite{Sycamore} is based on superconducting Xmon qubits~\cite{Kjaergaard20}, IonQ~\cite{IonQ} works with trapped ions~\cite{TrappedIon}, etc. IBM Quantum (IBM-Q) provides free cloud access to its quantum processors based on superconducting transmon qubits~\cite{IBMQ}. Superconducting qubits (transmon or Xmon type) are popular for running large-depth quantum circuits with many gate operations~\cite{Krantz19,Kjaergaard20}. This archetype has fewer gate times (i.e., the time required to perform one gate operation) than other archetypes~\cite{Comparison_gate_times}. But it suffers from less decoherence times~\cite{Comparison_gate_times, NotAllQubitsAreEqual}. 

IBM-Q~\cite{IBMQ} devices have progressed since their inception, in terms of gate fidelities and decoherence times as compared to their contemporaries. Less number of qubits and simple connectivity make them amenable to studying their noise through randomized benchmarking protocols such as SRB and URB. IBM quantum computers are programmed using \textit{Qiskit}~\cite{Qiskit}, which is a Python-based open-source library developed by IBM Q team to program and execute quantum circuits in simulators or on IBM-Q devices.
The decoherence times~\cite{NotAllQubitsAreEqual} are two constants that specify the quality of the qubits. The first constant is called $T1$ or the time constant associated with the average time for a qubit to return from the excited state to the ground state naturally. The second constant is called $T2$ or the time constant associated with the average time for the qubit to change its phase.

Attempts have been made to implement SRB and its variants on different quantum computing platforms and archetypes~\cite{Harper, Knill, three-qubit, cycle-benchmarking,IonQ}, however as stated earlier, variants of SRB do not quantify the coherence in the gate errors which is a more useful metric to analyze and benchmark gate implementation in various quantum processors. For example, Harper \textit{et al.}~\cite{Harper} have implemented the [4, 2, 2] code and performed RRB on the logical codespace of [4, 2, 2]. They found a significant decrease in the infidelity of the two-qubit gates.

\subsection{Motivations }
The motivation for this work is to find a way to implement the URB protocol in a real quantum device.  The direct implementation of the
URB protocol in a quantum device is not possible using current technologies, as it requires the preparation of
mixed states. This is the main motivation to propose a modification of the URB protocol namely, m-URB protocol.

The motivation for the proposal of Native gate URB (Ng-URB) is as follows.
\begin{enumerate}
    \item \textit{Firstly}, that the m-URB and SRB both are sensitive to the actual compilation of the gates in the gate set on a specific platform or archetype. This implies that the same gate set may provide different values of the figure of merit used, depending on the actual compilation of gates. Although the protocols inherently are platform or archetype-agnostic, the actual implementation then becomes device-dependent, essentially because of compilation. 
    \item \textit{Secondly}, to benchmark the performance of quantum computation on a specific platform or archetype, characterizations-induced is associated with individual native gates is desirable, as it is more informative about the overall operating characteristics of the processor.
\end{enumerate}
\subsection{Contributions}
The contributions of the paper are listed below.
\begin{enumerate}
    \item \textit{Firstly}, we simulate the URB protocol using Qiskit~\cite{Qiskit}, which is an open-source library to program and execute quantum programs in simulators or on actual IBM-Q devices. To the best of our knowledge, our work presents the first step in the direction of simulating benchmarking the coherence of noise induced during the execution of quantum programs on any quantum device. As compared to SRB and its other variants, the implementation of URB requires the preparation of mixed states. Preparing such states with full user-defined control is not possible today on any known platform. However, we propose a slight modification in the protocol, we call it modified URB (m-URB) protocol that enables one to practically implement the URB protocol on any quantum device.
    \item \textit{Secondly}, we simulated the effectiveness of our m-URB protocol, to gauge the coherence of different noise channels. We perform  experiments using m-URB protocol by simulating two well-studied noise channels; 

(a) the single-qubit depolarising channel and 

(b) the single-qubit bit-flip channel. 

\noindent We also validate our implementation by comparing it with the theoretical unitarity values of the URB protocol. Hence, we provide the proof of principle for our successful implementation of the m-URB protocol and its soundness.
\item \textit{Thirdly}, we propose another modification to the m-URB protocol, which we call the native gate URB (Ng-URB), in order to study the noise in the native gates, i.e., the gates where the quantum circuits are compiled and executed in the actual IBM-Q devices. We perform a procedure to interleave ideal and noisy gates to benchmark the simulated backend noise of two of the IBM-Q processors, Burlington device~\cite{IBMQ_Burlington} and Melbourne device~\cite{IBMQ_Melbourne} which are $5$-qubit and $15$-qubit processors respectively.
\item \textit{Finally}, using Ng-URB protocol we detect the presence of cross-talk errors~\cite{Cross_talk} which are correlated errors caused due to non-local and entangling gates such as CNOT gate. We find that both the IBM-Q processors have a noticeable but considerably low cross-talk when applying native CNOT gate, based on the unitarity values. However, there is a considerable difference in the unitarity of the CNOT gate for the $5$-qubit processor as compared to the $15$-qubit one. We hope that our research highlights the practical effectiveness of the Ng-URB protocol in successfully detecting the cross-talk between qubits when applying native gate operations irrespective of the platform.
\end{enumerate}

This paper is structured as follows. We first introduce key concepts related to URB in Sec.~\ref{Overview}. 
The details of our first protocol, i.e., m-URB protocol are discussed in Sec.~\ref{mURB} where in Sec.\ref{state_prep}, we describe the procedure to prepare single and two-qubit mixed states by classical post-processing; in Sec.~\ref{measure}, we show how to calculate the expectation of any Pauli observable; and in Sec.~\ref{derivation}, we present the theoretical calculation of unitarity for custom noise channels.
In Sec.~\ref{NativeURB}, we describe the details of our second protocol, i.e., Ng-URB protocol, whereas, in Sec.~\ref{theoretical_Ng_URB} and Sec~\ref{Implementation_Ng_URB}, we present the theoretical and implementation details of this protocol respectively. The simulation results of our protocols and related discussion are presented in Sec.~\ref{Simulation}. The results of m-URB for depolarising channel and bit-flip channel is shown in Sec.~\ref{Noise_sims}. The implementation details of Ng-URB protocol in 5-qubit and 15-qubit processors are provided in Sec.~\ref{NativeURBResults}. The analysis of the results of Ng-URB protocol and the detection of cross-talk error is presented in Sec.~\ref{discussion}. The comparison between the work of~\cite{Sheldon} and our work is presented in Sec.~\ref{Comparison with Sheldon}. Finally, the concluding remarks are presented in Sec.~\ref{Conclusion}.

\section{Overview of Unitarity Randomized benchmarking (URB)} \label{Overview}

In this section, we state the important concepts to understand the URB protocol, namely purity and unitarity~\cite{Wallman}. While purity is essentially the square of the Hilbert-Schmidt norm 
of a density operator, the unitarity of a noise channel is the figure of merit estimated by the URB protocol and is structurally similar to purity. Here, we define both purity and unitarity and highlight their similarity with each other. Next, we discuss the protocol in the literature to estimate unitarity experimentally. 
\subsection{Basic Definitions} First we will define purity and three structurally different but equivalent definitions of unitarity.
\subsubsection{Purity} Purity of a quantum state  expressed as a density operator $\rho$ is given by $\text{Tr} [ \rho^{\dagger}\rho] \in [0,1]$.
 Note that, $\text{Tr}[\rho^{\dagger}\rho] = 1$ if and only if $\rho$ is a pure state and it has a state vector representation in the Hilbert space $\mathcal{H}$ of dimension $d = 2^n$, where $n$ is the number of qubits.
 
\subsubsection{Unitarity} Let $\Lambda$ be a general \textit{quantum operation}~\cite{PreskillNotes} which defines the noise channel as a superoperator mapping the space of density operators
to itself. Note that for the sake of completeness, here we assume that noise channels are non-trace-increasing, and in particular, a noise channel can also be trace decreasing~\cite{Wallman}. A natural candidate to quantify the coherence of a noise channel, i.e., unitarity can be 
\begin{equation*}
\int d\psi \text{Tr}[\Lambda (\psi)^{\dagger}\Lambda (\psi)],
\end{equation*}
where the integral is with respect to uniform Haar measure over the space of all pure states $\ket{\psi} \in \mathcal{H}$. In other words, it is the purity of the output state after subjecting to the noise channel, averaged over all possible input pure states. Wallman \textit{et al.}~\cite{Wallman} argue that this definition of unitarity may not work for general quantum operations that are not completely positive trace preserving (CPTP) maps~\cite{PreskillNotes}. 

We mention here three structurally different but equivalent definitions of unitarity. 
\begin{enumerate}
    \item The first definition for unitarity as proposed by Wallman \textit{et al.} in~\cite[Eq. 4]{Wallman} is given by
\begin{equation} \label{unitarity1}
    u(\Lambda) = \frac{d}{d-1}\int d\psi 
 \text{Tr}[\Lambda^{'}(\psi)^{\dagger}\Lambda^{'}(\psi)],
\end{equation}
where $\Lambda^{'}(\psi) = \Lambda(\psi) - \frac{\text{Tr}[\Lambda(\psi)]}{\sqrt{d}}\mathbb{I}$. In other words, according to this definition, unitarity is the purity of the output state averaged over all possible input states when subjected to a modified noise channel which we get by subtracting the identity component of the channel from itself. 

\item Dirkse \textit{et al.}~\cite{Dirkse} propose the following second equivalent definition which is 
\begin{equation} \label{unitarity2}
    u(\mathcal{E}) = \frac{d}{d-1}\int d\psi \text{Tr}\Big\{\Big[\mathcal{E}\Big(\ket{\psi}\bra{\psi} - \frac{\mathbb{I}}{d}\Big)\Big]^{2}\Big\},
\end{equation}
where the integral is again with respect to uniform Haar measure over the space of pure states $\ket{\psi} \in \mathcal{H}$.
Here, $\mathcal{E}$ is a CPTP map and hence they differentiate it from a quantum operation by calling it a \textit{quantum channel}.
It is noteworthy, that in order to convert unitarity as a normalized metric the constant factor $\frac{d}{d-1}$ has been included. 

\item The third equivalent definition of unitarity as proposed in~\cite{Wallman, Dirkse} and which we heavily rely in this paper is: 
\begin{equation} \label{unitarity3}
    u(\mathcal{E}) = \frac{1}{d^{2} - 1}\sum\limits_{\sigma, \tau \in \mathbb{P}^{*}}\Big\{\text{Tr}[\tau\mathcal{E}(\sigma)]\Big\}^{2}.
\end{equation}
Here, $\mathbb{P}^{*}$ is the normalized Pauli group, excluding the normalized identity (the normalization is with respect to the Hilbert-Schmidt norm). This is the form of unitarity that we will use in its estimation protocol through URB as suggested in~\cite{Dirkse}. We discuss the equivalence of this form with the aforementioned forms in Sec.~\ref{Significance}.
\end{enumerate}

\subsection{Description of the protocol}
Now we formally state the URB protocol followed by a discussion on the user-defined parameters used in it. The authors in ~\cite{Wallman, Dirkse} propose two equivalent implementations of URB known as the \textit{single copy} and \textit{two copy implementation}. We remark here, that we follow \textit{single copy implementation} in our experiments, and therefore, we describe here only the former implementation. The \textit{two copy implementation} and its equivalence and comparison with \textit{single copy implementation} can be found in~\cite{Dirkse}.

\RestyleAlgo{boxruled}
\begin{algorithm}
\caption{Pseudocode for URB Protocol for single copy implementation}\label{Algo:1}
Gate set $\mathbb{G}$, sequence lengths (depths) $\mathbb{M}$, Number of iterations (or number of random sequences) $\text{N}_{m}$ for every depth $m \in \mathbb{M}$. Let $d = 2^n$, where $n$ is the number of qubits.
\begin{algorithmic}[1]
\Procedure{URB, $\mathbb{M}$, $ \lbrace \text{N}_{m} \rbrace$}{}
\State \textbf{for all} depth $m \in \mathbb{M}$
\State \hspace{0.2cm} \textbf{repeat} $\text{N}_{m}$ times 
\State \hspace{0.4cm} Sample a sequence of $m$ gates $\mathcal{G}_{j_{1}},\mathcal{G}_{j_{2}},...,\mathcal{G}_{j_{m}}$
uniformly at random from $\mathbb{G};$
\State \hspace{0.4cm} Compose the sequence $\mathcal{G}_{\textbf{j}} = \mathcal{G}_{j_{m}}...\mathcal{G}_{j_{2}}\mathcal{G}_{j_{1}};$ 
\State \hspace{0.6cm} \textbf{for all} non-identity Pauli's \textit{P}, \textit{Q} 
\State \hspace{0.8cm} Prepare $\rho^{(P)} = \frac{I+P}{d}$ and $\hat{\rho}^{(P)} = \frac{I-P}{d}$;
\State \hspace{0.8cm} Apply $\mathcal{G}_{\textbf{j}}$ to each state;
\State \hspace{0.8cm} Measure $E^{(Q)} = Q$ large number of times;
\State \hspace{0.8cm} Estimate the shifted purity for a sequence,  $$ \quad q_{\text{\textbf{j}}} = \frac{1}{d^2 - 1} \sum\limits_{\text{{\tiny P,Q $\neq$ I}}}\{ \text{\text{Tr}}[E^{(Q)}\mathcal{G}_{\textbf{j}}\rho^{(P)}]-\text{\text{Tr}}[E^{(Q)}\mathcal{G}_{\textbf{j}}\hat{\rho}^{(P)}]\}^{2};$$
\State \hspace{0.2cm} Average over all sequences $\Tilde{q}_{m} = \frac{1}{\text{\textbf{N}}_{m}}\sum\limits_{\textbf{j}} q_{\text{\textbf{j}}}$;
\State Fit exponential curve $\Tilde{q}_{m} = Bu^{m-1}$, where $u$ is unitarity and $B$ is a constant absorbing SPAM.
\EndProcedure
\end{algorithmic}
\end{algorithm}

 Below, we discuss the parameters used in the protocol mentioned in the algorithm above.

\subsubsection{Gate set ($\mathbb{G}$)} The set of gates from which random gates are sampled and applied to the initial state. The motivation for defining a gate set for choosing random gates is similar in spirit to that of averaging the noise channel over all possible input states, as described in SRB protocol. We remark here, that in~\cite{Wallman, Dirkse} as well as in this paper, $\mathcal{E}$, i.e., the noise channel is gate-independent and hence the actual noise map associated with a gate in the gate set is close to Haar averaged noise over all unitary gates concerning Hilbert-Schmidt norm. We chose our gate set to be the Clifford group (i.e., single and two-qubit Clifford groups for estimating unitarity of noise induced in single and two-qubit gates respectively) only for the experiments with simulated custom noise models.

 \subsubsection{Sequence lengths (depths)($\mathbb{M}$)} An array of several gates to be sampled from the gate set $\mathbb{G}$ for empirically estimating the unitarity. For each depth value of $m$ in the array $\mathbb{M}$, the depth many gates are sampled and applied to the initial state which results in the composition of the averaged noise channel over the gate set with itself depth many times and being applied to the initial state. Ideally, it is recommended to use many different values for the depths, ranging from low number of gates to high. This has the effect of lowering the variance in the estimation of unitarity, obtained as a result of fitting the exponential curve mentioned in Algorithm~\ref{Algo:1}. We also remark, that in the cases where the noise itself is largely incoherent, large depth values may lead to low signal strengths (i.e. the estimated shifted purities per iteration mentioned in~\ref{Algo:1}) which leads to an overestimation of unitarity, as we discuss in Sec.~\ref{Noise_sims}.

\subsubsection{Number of iterations ($\text{N}_{m}$)} For each depth $m \in \mathbb{M}$ and in each iteration, shifted purity is calculated with a different sequence of length $m$ sampled from $\mathbb{G}$. The shifted purities are averaged over $\text{N}_{m}$ many iterations, to get the empirical average shifted purity $\Tilde{q}_{m}$ for a given depth $m$. Again, it is recommended to choose high enough $\text{N}_{m}$ so that between sequence variance $\mathbb{V}[q_{\textbf{j}}]$ is minimized. Dirkse \textit{et al.}~\cite{Dirkse} gave a theoretical lower bound on $\text{N}_{m}$ for given confidence parameters $\epsilon \text{ and } \delta$, where $\epsilon$ can be treated as the accuracy or the deviation of the empirical average $\Tilde{q}_{m}$ from $\mathbb{E}[q_{\textbf{j}}]$ and $\delta$ as upper bound on the probability that $\Tilde{q}_{m}$ is `more than $\epsilon$ away' from $\mathbb{E}[q_{\textbf{j}}]$. The relation between $\epsilon \text{ and } \delta$ can be mathematically expressed by the concentration inequality,
 \begin{equation}
 \text{Pr}[|q_{\textbf{j}}-\mathbb{E}[q_{\textbf{j}}]|\geq \epsilon]\leq\delta.    
 \end{equation}
  The theoretical bound on $\text{N}_{m}$ is typically in the range of $\sim 200$ for suitable confidence parameters. However, we remark that since our work is completely based on simulations, a smaller number of iterations suffice to produce results with desirable confidence parameters and with low variance.

\subsection{Discussion}
This concludes a detailed discussion on the URB protocol. Here, we would like to mention two comments.
\begin{enumerate}
    \item \textit{Firstly}, that Wallman \textit{et al.} derive that the unitarity, similar to average gate fidelity, decays exponentially with increasing the number of gates in the gate sequence applied and is given by, $\mathbb{E}[q_{\textbf{j}}] = A + Bu^{m-1}$~\cite{Wallman}. Dirkse \textit{et al.} propose a modification, to prepare two separate initial states $\rho \text{ and } \hat{\rho}$ and apply the same sequence $\mathcal{G}_{\textbf{j}}$ to each of the states simultaneously, so that we effectively work with a traceless density operator $\Tilde{\rho}:=\frac{1}{2}(\rho - \hat{\rho})$ that changes the form of the fit curve to $\Tilde{q}_{m} = Bu^{m-1}$~\cite{Dirkse}. This is easier to fit, since by taking the logarithm of the empirical average purities and plotting them against depths we can reduce it to a linear regression problem with the logarithm of the unitarity as a fit parameter of the regression.

\item \textit{Secondly}, the mathematical form of the two initial states $\rho^{(P)} \text{ and } \hat{\rho}^{(P)}$ to be prepared as mentioned in the single copy URB implementation in Algorithm~\ref{Algo:1}, is advocated in~\cite{Dirkse}. The authors argue that the SPAM absorbing constant B in the fit model $\Tilde{q}_{m} = Bu^{m-1}$ has the property that $|B| \leq 1$~\cite[Lemma. 11]{Dirkse}. Since we would ideally want to maximize the signal strength, for the channels which are vastly incoherent (unitarity less than say, $0.5$) we would like to choose our initial states and measurement operators so that  $B = 1$ and the choice of $\rho^{(P)} = \frac{I+P}{d}, \hat{\rho}^{(P)} = \frac{I-P}{d} \text{ with } E^{(Q)} = Q$ (for \textit{single copy implementation}) is the two-qubit choice for which $B = 1$. Another reason for this choice is the fact that Pauli operators are natural observables, for which we can easily calculate the expectation values (i.e., the trace terms inside the expression for shifted purity in Algorithm~\ref{Algo:1}). We remark, that though the initial states proposed are mixed states for $n \geq 2$ qubits, the mixed states required to be prepared can be easily expressed as the convex sum of pure states which in turn can be easily prepared in actual implementation. We discuss this remark in more detail in Sec.~\ref{mURB}.
\end{enumerate}

\section{Details of our Proposed m-URB protocol} \label{mURB}
The brief overview of this section is as follows.
\begin{enumerate}
    \item  \textit{First}, we describe the implementation of our proposed m-URB protocol to characterize the noise associated with multi-qubit gates. Although our protocol works for any $n$-qubit gate, we will only consider $n \in \lbrace 1, 2 \rbrace$ since, it is known that only single and two qubit gates are sufficient to realize universal quantum computation. We specifically focus on step 6 of Algorithm~\ref{Algo:1} which in general requires the states $\rho^{(P)}$ and $\hat{\rho}^{(P)}$ to be mixed states. Below, we provide a modification to this step in Sec.~\ref{state_prep} to make this step practically feasible.

\item \textit{Next} in Sec.~\ref{derivation} we derive the theoretical unitarity of some well-known noise channels, namely, (a) (completely) depolarising channel and (b) the single qubit bit-flip channel, using the definition of unitarity as discussed in Eq.~\eqref{unitarity3}. The results in Sec.~\ref{derivation} will be used to compare with our unitarity estimates of the same noise channels simulated in Qiskit. These noise channels are used to simulate the gate noise for a gate set. We remark, that the agreement between theoretical unitarities and the estimated unitarities obtained via the m-URB protocol for the simulated noise channels highlights the proof of concept for our m-URB protocol.
\end{enumerate}
Note that in the m-URB protocol, we use the Clifford group as our gate set and implement the protocol for single and two-qubit Clifford unitaries. For implementing a Clifford gate, we use the library QuaEc: Quantum Error Correction Analysis in Python~\cite{QuaEc}, for generating the single and two-qubit Clifford group for the purposes of randomly sampling a Clifford unitary and decomposing a Clifford gate into some basic well-known gates namely, \textit{I, X, Y, Z, H, T, controlled-$X$, controlled-$Z$} and \textit{SWAP} gates. Using the library, we first extract the QASM (Quantum Assembly Language) string~\cite{QASM} specifying the Clifford gate decomposition and then we implement the QASM instruction manually in Qiskit. We also use the Clifford group as our gate set for empirically estimating the unitarity of the two simulated noise models as discussed in Sec.~\ref{derivation}.


\subsection{Single and two-qubit state preparation} \label{state_prep}
According to the standard approach used to empirically calculate unitarity mentioned in~\ref{Algo:1}, Step 6 of the algorithm requires preparation of two initial states $\rho^{(P)} = \frac{(I+P)}{d}$ and $\hat{\rho}^{( P)} = \frac{(I+P)}{d}$, where $P$ is a non-identity Pauli gate over $n$ qubits.

 Notice that the states $\rho^{(P)}$ and $\hat{\rho}^{( P)}$ are the density matrices corresponding to the positive and negative eigenstates of the Pauli operator $P$. This is because the expectation value of the Pauli operator $P$ in states $\rho^{(P)}$ and $\hat{\rho}^{( P)}$ is `+1' and `-1' respectively i.e.
 \begin{equation}
    \text{Tr}(P(I+P)/d) = 1
 \end{equation} and 
 \begin{equation}
     \text{Tr}(P(I-P)/d) = -1
 \end{equation}
 For instance, for single qubit case i.e. $n = 1$, there are only three choices for $P$ namely $X$, $Y$, $Z$ gates (i.e. the Pauli operators $X$, $Y$ and $Z$ respectively). If $P = X$ then $(I+X)/2 = \ket{+}\bra{+}$ and $(I-X)/2 = \ket{-}\bra{-}$, where $\ket{+}$ and $\ket{-}$ are respectively the positive and negative eigenstates of Pauli $X$ operator. Therefore in the single qubit case, for each choice of $P$ we initialize two separate circuits with the positive and negative eigenstates as the initial states. Next, we apply the same randomly generated sequence of Clifford unitaries to both of the circuits. However for $n > 1$ case i.e. when $d = 2^{n} > 2$, both $\rho^{(P)}$ and $\hat{\rho}^{(P)}$ are mixed state density matrices as, 
 \begin{equation}
     \text{\text{Tr}} \left((I+P)(I+P)/d^2 \right) = \text{\text{Tr}} \left(2I/d^2 \right) = (2/d) < 1.
 \end{equation} Similarly,
 \begin{equation}
    \text{\text{Tr}}((I-P)(I-P)/d^2) = \text{\text{Tr}}(2I/d^2) = (2/d) < 1 
 \end{equation}
where we have used the fact that $P$ is a normalized non-identity Pauli operator and that $P$ is traceless.

Therefore we can say that in general the states $\rho^{(P)}$ and $\hat{\rho}^{(P)}$ are mixed states for $n > 1$. To implement the URB protocol as mentioned in Algorithm~\ref{Algo:1} for a two-qubit case, one way is to directly prepare the input mixed states which is not supported currently in any available platform or archetype. Here, we emphasize that although mixed states are routinely created in currently available quantum processors, they are mostly created as a result of incoherent noise (non-unitary operation) in the computation. This is not a desirable means to create mixed states since there is no `user-defined control' to accurately create any arbitrary mixed state. It is this `user-defined control' (for instance, using microwave pulse shaping for superconducting qubits) to creates states with arbitrary density matrices, which is not possible today to the best of our knowledge. Therefore, we opt for classical post-processing to simulate the statistics of a mixed state using the statistics obtained for the pure states present in its convex decomposition. In particular, for two-qubit state preparation, we first decompose the two-qubit mixed states into the convex sum of pure state density matrices. For instance, if $P = X \otimes X$ then $\rho^{(P)}$ can be written as 
\begin{equation} \label{rho_p}
\begin{aligned}
\frac{I+X\otimes X}{4} =\frac{\ket{\psi^{+}}\bra{\psi^{+}} + \ket{\phi^{+}}\bra{\phi^{+}}}{2},
\end{aligned}
\end{equation}
and $\hat{\rho}^{(P)}$ can be written as
\frenchspacing
 \medmuskip=0mu
\thinmuskip=0mu
\thickmuskip=0mu 
\begin{equation} \label{hat_rho_p}
\begin{aligned}
\frac{I-X\otimes X}{4} = \frac{\ket{\psi^{-}}\bra{\psi^{-}} + \ket{\phi^{-}}\bra{\phi^{-}}}{2}, 
  \end{aligned}
\end{equation}
\frenchspacing
 \medmuskip=2mu
\thinmuskip=2mu
\thickmuskip=2mu 
where $\ket{\psi^{\pm}} = \frac{\ket{00} \pm \ket{11}}{\sqrt{2}}$, $\ket{\phi^{\pm}} = \frac{\ket{01} \pm \ket{10}}{\sqrt{2}}$.

One can show that, for each non-identity Pauli gate $P$, both $\rho^{(P)}$ and $\hat{\rho}^{(P)}$ can be written as the sum of two pure state density matrices as demonstrated for $P = X$ above. Therefore as per our m-URB protocol for the two-qubit case, we initialize four circuits per choice of $P$, with a pair of circuits initialized with the corresponding pure states in the decomposition of each of the states $\rho^{(P)}$ and $\hat{\rho}^{(P)}$. For example, for the case of $P = X \otimes X$, we first initialize two circuits with initial states prepared in the Bell states $\ket{\psi^{+}}$ and $\ket{\phi^{+}}$ and apply the same random sequence of Clifford gates to each of the circuits. Similarly, we then initialize another pair of circuits with Bell states $\ket{\psi^{-}}$ and $\ket{\phi^{-}}$ as initial states and apply the same random sequence of Clifford gates as above to each of them. We note, that the factor of $0.5$ in the density matrix representation of $\rho^{(P)}$ and $\hat{\rho}^{(P)}$ given by Eqs.~\ref{rho_p} and~\ref{hat_rho_p} is taken into account in the measurement step as discussed below.

\subsection{Calculating the expectation of Pauli observable}
\label{measure}
For completeness, we again state here the expression for shifted purity mentioned in Step 9 of Algorithm~\ref{Algo:1}:

\begin{equation} \label{shifted_purity}
   q_{\text{\textbf{j}}} =  \dfrac{1}{d^2 - 1} \sum\limits_{\text{P,Q $\neq$ I}} \left\lbrace \text{\text{Tr}} \left[E^{(Q)}\mathcal{G}_{\textbf{j}}\rho^{(P)} \right] -\text{\text{Tr}} \left[ E^{(Q)}\mathcal{G}_{\textbf{j}}\hat{\rho}^{(P)}\right] \right \rbrace^{2}.
\end{equation}

Here, $E^{(Q)} = Q$, where $Q$ is any non-identity normalized Pauli observable. Notice the term in the summation is nothing but the square of the difference in the expectation values of the Pauli observable $Q$, in the quantum states obtained after the application of the random sequence of Clifford gates to the input states $\rho^{(P)}$ and $\hat{\rho}^{(P)}$. 

In the m-URB protocol that we propose, for all circuits initialized in the previous state preparation step in Sec.~\ref{state_prep}, we calculate the expectation value of $Q$ after applying the same random sequence of Clifford gates. Note that, since any Pauli observable has only two eigenspaces corresponding to its two possible eigenvalues i.e. +1 and -1, we can essentially perform a positive operator-valued measurement (POVM). Suppose the POVM element corresponding to positive eigenspace of $Q$ is $M$, then we can write 
\begin{equation}
    Q = 2M - I
\end{equation} so that 
\begin{equation} \label{povm_eqn}
    \text{\text{Tr}}[Q\rho] = 2\text{\text{Tr}}[M\rho] - 1
\end{equation}
where $\text{\text{Tr}}[M\rho]$ is nothing but the probability that the state $\rho$ collapses to positive eigenspace of Q. Since in most of the quantum processors the measurement can only be done in the computational basis, therefore we rotate the basis by a change of basis transformation. First we define $\Tilde{Z}$ as follows
\begin{equation}
    \Tilde{Z} = Z \text{ if Q $\in$ \{$X$, $Y$, $Z$\}} 
\end{equation}
and 
\begin{equation}
    \Tilde{Z} = \begin{cases}
    Z \otimes I \hspace{0.2cm} \text{when} \hspace{0.2cm} Q \in \lbrace X \otimes I, X \otimes Z, Y \otimes I, Y \otimes Z \rbrace,\\
     I \otimes Z \hspace{0.2cm} \text{when} \hspace{0.2cm} Q \in \lbrace I \otimes X, Z \otimes Z, I \otimes Y, Z \otimes Y \rbrace, \\
      Z \otimes Z \hspace{0.2cm} \text{when} \hspace{0.2cm} Q \in \lbrace X \otimes X, Y \otimes Y, X \otimes Y, Y \otimes X  \rbrace.
    \end{cases}
\end{equation}
Now we can rotate the basis as follows

\begin{equation}
    \text{\text{Tr}} \left[ Q\rho \right] = \text{\text{Tr}} \left[ U^{\dagger} \Tilde{Z} U\rho \right] = \text{\text{Tr}}\left[\Tilde{Z} U\rho U^{\dagger}\right],
\end{equation}
where $U$ is a unitary operator that satisfies $U^{\dagger} \Tilde{Z} U = Q$ and the second equality corresponds to the change of basis transformation to $\Tilde{Z}$ basis.


Note that for Pauli $Z$ gate, the POVM element corresponding to the positive eigenspace of $Z$ is just $\ket{0}\bra{0}$. Therefore Eq.~\ref{povm_eqn} essentially equates to calculating the probability that the final state collapses to $\ket{0}$ state in the single qubit implementation. On the other hand for two-qubit implementation of the protocol, depending on the observable $Q$, we consider only the subset of computational basis states spanning the positive eigenspace of Pauli observable $\Tilde{Z}$ as mentioned above and calculate the empirical probability of the state collapsing into one of these basis states. For instance, suppose $Q = X$ then $U = H$ (Hadamard gate) as it is known that $HZH = X$. Hence for calculating the expectation of $Q$, we apply an additional Hadamard gate to each of the circuits and then estimate the probability that the quantum register in each circuit collapses in $\ket{0}$. Let $p_{1}$ and $p_{2}$ be the corresponding probabilities of $\ket{0}$ obtained for circuits initialized with $\rho^{(P)}$ and $\hat{\rho}^{(p)}$ respectively, then $\text{\text{Tr}}[X\mathcal{G}_{\textbf{j}}\rho^{(P)}] = 2p_{1} - 1$ and $\text{\text{Tr}}[X\mathcal{G}_{\textbf{j}}\hat{\rho}^{(P)}] = 2p_{2} - 1$ for some choice of $P$ and sequence $\mathcal{G}_{\textbf{j}}$. 

As mentioned before in the single qubit case, both $\rho^{(P)}$ and $\hat{\rho}^{(p)}$ are pure states that can be directly prepared. Hence, for each observable $Q$ we calculate the expectation of $Q$ for the two circuits corresponding to $\rho^{(P)}$ and $\hat{\rho}^{(p)}$ and for every choice of observable $P$. However, in the two-qubit case since we prepare two circuits for each of the states $\rho^{(P)}$ and $\hat{\rho}^{(p)}$ we calculate the expectation of $Q$ for four circuits and for each choice of observable $P$. Finally, in the two-qubit case, we multiply each expectation value obtained, by the factor 0.5, as each of $\rho^{(P)}$ and $\hat{\rho}^{(p)}$ are equal mixtures of their constituent pure states for each $P$, as shown in the example for $P = X\otimes X$ above.
In other words, suppose the probability that the quantum register collapses to positive eigenspace of $Q = X \otimes X$ in each of the circuits initialized with pure states $\ket{\psi^{+}}$, $\ket{\phi^{+}}$, $\ket{\psi^{-}}$ and $\ket{\phi^{-}}$ and for the case of $P = X \otimes X$ be, $p_{1}$, $p_{2}$, $p_{3}$, and $p_{4}$ respectively, then
\begin{equation}
    \begin{aligned}
    \text{\text{Tr}}\Big[X\otimes X\mathcal{G}_{\textbf{j}}\rho^{(P)}\Big] = & \frac{1}{2}\Big(\text{\text{Tr}}\Big[X\otimes X\mathcal{G}_{\textbf{j}}\ket{\psi^{+}}\bra{\psi^{+}}\Big] \\ & +  \text{\text{Tr}}\Big[X\otimes X\mathcal{G}_{\textbf{j}}\ket{\phi^{+}}\bra{\phi^{+}}\Big]\Big) \\
    =&  \frac{1}{2}(2p_{1} - 1 + 2p_{2} -1),
    \end{aligned}
\end{equation}
and 
\begin{equation}
    \begin{aligned}
    \text{\text{Tr}}\Big[X\otimes X\mathcal{G}_{\textbf{j}}\hat{\rho}^{(P)}\Big] = & \frac{1}{2}\Big(\text{\text{Tr}}\Big[X\otimes X\mathcal{G}_{\textbf{j}}\ket{\psi^{-}}\bra{\psi^{-}}\Big] \\ & +  \text{\text{Tr}}\Big[X\otimes X\mathcal{G}_{\textbf{j}}\ket{\phi^{-}}\bra{\phi^{-}}\Big]\Big) \\ = &  \frac{1}{2}(2p_{3} - 1 + 2p_{4} -1).
    \end{aligned}
\end{equation}
The expectation values are then plugged into Eq.~\ref{shifted_purity} to get the shifted purity. After calculating the shifted purity, we further divide by a factor of 4, to match with the theoretical expression of unitarity, which comes out to be one-fourth the shifted purity as mentioned in Sec.~\ref{Significance}. We remark that Dirkse \textit{et al.} in~\cite{Dirkse} did not mention dividing by 4 in order to keep the signal strength, i.e. the shifted purity values for each iteration high enough even in the case of extreme noise channels which have very low unitarity. However, we adopt to divide by a factor of 4 in our implementation. Since IBM-Q devices are sufficiently robust, the noise induced by the gate operations lies in high unitarity range and hence we do not gain much by keeping the factor of 4. 

We also remark that following the suggestion of Dirkse \textit{et al.} in~\cite{Dirkse}, we also run the same experiment i.e. a given choice of $P$, $Q$ and $\mathcal{G}_{\text{\textbf{j}}}$ multiple numbers of times, to take into account for the within sequence variance in the estimation of unitarity which we call `the number of samples'.

\subsection{Theoretical calculation of unitarity for custom noise channels} \label{derivation}
We now derive the theoretical unitarity for two well-known noise models. First, we consider the single qubit (completely) depolarising channel with arbitrary depolarising parameter $0 < p < 1$. Next, we consider the single qubit bit-flip channel which is a special case of the general mixed unitary channel. We remark that although we consider both the noise channels for a single qubit, the derivations can be easily generalized to multi-qubit cases and for other variations of the noise channels considered here. Later we implement the m-URB protocol, with the above-mentioned noise channels simulating the noise in the single-qubit Clifford gates as discussed in Sec.~\ref{Noise_sims}.

\subsubsection{Unitarity of a depolarizing noise map} 
\label{theoretical_dep}
Let $\mathcal{E}$ be the superoperator corresponding to a (completely) depolarising noise~\cite{Nielsenbook} over the Hilbert space $\mathbb{C}^{2^{n}}$, where $n$ is the number of qubits and with depolarising parameter $p$. The depolarising parameter can be considered as the probability that a given quantum state $\rho$ is unchanged after the noise is applied. Mathematically, the depolarising channel is written as

\begin{equation} \label{depolarising_channel}
    \mathcal{E}(\rho) = p\rho + (1-p)\text{\text{Tr}}[\rho]\frac{\mathbb{I}}{2^{n}}.
\end{equation}

 Here, $\mathbb{I}$ is the $2^n \times 2^n$ identity matrix. Using Eq.~\eqref{unitarity3} and the fact that $\text{\text{Tr}}[\sigma] = 0$, $\forall\sigma \in \mathbb{P}^{*}$, theoretical unitarity of depolarising channel defined in Eq.~\eqref{depolarising_channel} is given by, 

\begin{equation} \label{unitarity_depolarising_channel}
\begin{aligned}
u(\mathcal{E}) =& \frac{1}{4^{n} - 1}\sum\limits_{\sigma,\tau \in \mathbb{P^{*}}}\Big\{\text{\text{Tr}}\Big[p\tau\sigma+(1-p)\text{\text{Tr}}[\sigma]\frac{\tau}{2^{n}}\Big]\Big\}^{2}\\
=& \frac{1}{4^{n} - 1}\sum\limits_{\sigma,\tau \in \mathbb{P^{*}}}\Big\{p\text{\text{Tr}}[\tau\sigma]\Big\}^{2}
= \frac{1}{4^{n} - 1}\sum\limits_{\sigma \in \mathbb{P}^{*}} p^{2} = p^{2},
\end{aligned}
\end{equation}
where we have used the fact that $\sigma, \tau \in \mathbb{P}^{*}$ are orthonormal with respect to Hilbert-Schmidt inner product and $|\mathbb{P}^{*}| = 4^{n} - 1$.

\subsubsection{Unitarity of Bit-flip channel}
\label{theoretical_bit_flip}
A mixed unitary noise map is defined as the convex sum of unitary noise channels written below~\cite{Nielsenbook}
\begin{equation} \label{mixed_unitary_noise_map}
    \mathcal{E}(\rho) = \sum\limits_{k = 0}\limits^{n-1}p_{k}U_{k}\rho U_{k}^{\dagger}.
\end{equation}
Here, $\{U_{0}, U_{1}, U_{2}, \ldots ,U_{n-1}\}$ is the set of unitary operators (which can also include identity operator) and $\{p_{k}\}_{k=0}^{n-1}$ is the probability that the quantum state $\rho$ is subjected to the unitary noise $U_{k}\rho U_{k}^{\dagger}$ associated with unitary $U_{k}$. Note that the mixed unitary channel or noise map is in general incoherent in nature, unless $p_{k} \neq 0$ for some $k$ and $p_{m} = 0$ $ \forall m \neq k$.

One particular case of the mixed unitary noise channel which is widely studied with respect to quantum error correction literature, is the bit-flip channel~\cite{Nielsenbook}. For a single qubit register, it is represented as,

\begin{equation}\label{bit-flip}
    \mathcal{E}(\rho) = p\rho + (1-p)X\rho X.
\end{equation}

Intuitively, the action of the channel is such that with probability $p$, the qubit is left unchanged (i.e. applying identity channel), and with probability $1-p$, Pauli X is applied to the qubit. Although, we consider only the single qubit bit-flip channel, a generalization for multi-qubits is possible, for instance applying Pauli X to a subset of qubits with some probability. However, in the multi-qubit case, we have to consider the independence (dependence) of errors in the qubits i.e. whether the bit flips occur simultaneously for all the qubits or for any subset of them. For this reason, the computation of unitarity is more involved in the multi-qubit case and hence, we stick to the simplest case of single-qubit bit-flip channel for demonstration purposes.

Now we derive the theoretical unitarity of the bit flip channel defined in Eq.~\eqref{bit-flip} as follows.


\begin{equation}\label{unitarity_bit_flip}
\begin{aligned}
u(\mathcal{E}) =& \frac{1}{4^n-1}\sum\limits_{\sigma,\tau \in \mathbb{P}^{*}}\Big\{\text{Tr}\Big[\tau(p\sigma + (1-p)X \sigma X)\Big]\Big\}^{2}\\
=& \frac{8p^{2} - 8p + 3}{3}.
 \end{aligned}
\end{equation}

Here, $\Tilde{X}$ is the normalized Pauli X in $\mathbb{P}^{*}$ i.e. the normalized Pauli group for a single qubit.
We also use the fact that for a single qubit, $X\omega X = -\omega \hspace{0.2cm} \forall \omega \in \mathbb{P}^{*}$.

\section{Details of our proposed Ng-URB protocol} \label{NativeURB}
In this section, we modify the m-URB protocol and propose an alteration, to benchmark the noise associated with the native gates of an actual processor which we call \textit{Native gate URB} or \textit{Ng-URB}. 

\begin{enumerate}
    \item \textit{First}, we discuss the modifications made to the m-URB protocol in Sec.~\ref{theoretical_Ng_URB} along with a formal description of the protocol given in Algorithm~\ref{Algo:2}.

\item \textit{Next}, we explain the implementation details of the Ng-URB protocol in Sec.~\ref{Implementation_Ng_URB} to characterize the noise in the native gates of two of the IBM-Q processors, 5-qubit~\cite{IBMQ_Burlington} and 15-qubit~\cite{IBMQ_Melbourne}. The results of our implementation are documented in Sec.~\ref{NativeURBResults}, on the basis of which we comment on the extent of cross-talk in the two IBM processors.  
\end{enumerate}
\subsection{Motivation and Theoretical details} 
\label{theoretical_Ng_URB}

Most of the randomized benchmarking protocols (as mentioned in Sec.~\ref{Introduction}) quantify the performance of quantum computation by estimating some metric. Moreover, the protocols are designed to benchmark the average performance of gates in a gate set, which in most cases must satisfy certain properties such as the gate set must be a unitary 2-design. We propose that it is more practically insightful to benchmark the performance of the individual native gates of an actual processor to benchmark the quality of quantum computation that can be performed on a given device. We also note that Proctor \textit{et al.} have also argued in~\cite{Proctor} that the SRB protocol cannot provide the actual gate fidelities of the native gates provided by a platform or archetype, which ultimately defines the quality of computation on any actual device. In order to benchmark the performance of native gates, we introduce Native gate URB which to measure the coherence of the noise induced in the operation of native gates. Furthermore, we hope that studying the unitarity of a native gate can also help in understanding the physical sources of the noise in the actual implementation of the native gate in a given platform or archetype, thereby helping in the manufacturing of high-performance quantum processors.

`Native gates' are the unitary operations which have pre-defined calibrated instructions readily available, specific to a given quantum processor or platform. It is known that using a finite set of gates one can achieve universal quantum computation for example, the set of Clifford gates together with the T gate is known to be universal for quantum computation. For running general quantum circuits, each gate operation requested by the user is ultimately decomposed into these native gates before the actual execution of the quantum circuit. This process of decomposing each gate operation into a sequence of gates from the native gate set is also known as `transpiling'. 

When implementing SRB-type protocol on an actual device, the average gate infidelity is sensitive to the actual translation of the gates present in the circuit into the native gates available on a specific platform or archetype. This implies that even if the same random gate sequence is applied to two different transpired circuits, one may get different values for the infidelity. Therefore, although the SRB-type protocols are inherently platform or archetype-agnostic, the actual implementation of the protocol on a quantum processor depends on the quality of native gates and the efficient transpilation of the circuit. This means that the depth of the decomposed circuit should be as small as possible without redundant gates.

Motivated by the shortcomings of SRB-type protocols discussed above, we propose Ng-URB protocol. Firstly the protocol is platform agnostic. Secondly, it is transpilation agnostic and moreover, it directly computes the performance of the native gates individually instead of a gate set. Therefore, Ng-URB has characteristics of both IRB and DRB protocols described in Sec.~\ref{Introduction}. It benchmarks the unitarity of the noise associated with a single gate (as in IRB) and that gate belongs to the set of native gates pertaining to a processor or platform (as in DRB).  

\subsubsection{Modifications in the m-URB protocol}
Now we propose the following two modifications in the m-URB protocol to obtain the Ng-URB protocol.
\begin{enumerate}
    \item We consider no gate set and therefore no random sampling of gates is required. Note that sampling was required in SRB-type protocols to approximate the average noise channel for a gate set. Moreover, in the URB protocol mentioned in Algorithm~\ref{Algo:1} the use of a gate set is simply to calculate the average unitarity over the gate set. On the contrary, here we are interested in the performance of a single native gate. Since in SRB-type protocols the figure of merit is the average infidelity of the noise channel, where the average is taken with respect to all possible initial states (or equivalently the Haar Measure) a gate set is always required to `approximately' estimate the average. The reason that we can omit sampling from a gate set to calculate the unitarity of the noise associated with a single native gate is because of how unitarity is defined. 
    
   To make intuitive sense of why Ng-URB protocol does not require a gate set, we can interpret the superoperator $\mathcal{E}$ in the definition of unitarity in Eq.~\eqref{unitarity3} as the exact noise channel associated with a given native gate instead of the average noise over the gate set. The actual native gate operation can be thought of as a composition of two channels, the unitary channel which is the ideal native gate operation, and the noise channel associated with the native gate. Wallman \textit{et al.}~\cite[Proposition 7]{Wallman} show that the unitarity of a noise channel is unaffected if it is composed of any unitary channel. Therefore, instead of applying random sequences of gates from a gate set, we apply the native gate depth many times. Moreover, the unitarity can be interpreted as the Hilbert-Schmidt norm of the unital part of the superoperator $\mathcal{E}$ up to a constant factor (see Appendix~\ref{Significance}). In the URB protocol mentioned in Algorithm~\ref{Algo:1} we essentially estimate this trace norm at each depth value and then fit an exponential curve to the shifted purities vs. depth plot to extract the `estimated' unitarity. As evident from the discussion above, note that conceptually we don't require sampling from a gate set for the estimation protocol unless we specifically want to estimate `average unitarity' over a gate set. Here we remark that, although we have tried to give an intuitive explanation for the soundness of the Ng-URB protocol, a precise mathematical proof is a subject of future research.
    
    \item In each iteration and each depth $d$ of the circuit, a single (ideal) identity gate is applied, (for the two-qubit version, we apply $I \otimes I$) followed by the native gate in consideration and this is repeated $d$ times. This constitutes a sequence in the modified protocol. The identity gate is an ideal gate and the noise is imparted in computation solely by the actual native gate. The identity gate is interleaved because, IBM-Qiskit tries to optimize the circuit at run-time, by removing unnecessary repetition of gates (such as applying CNOT twice, which is just the identity gate), although we comment that, this has no effect on the unitarity calculated, as justified in~\cite[Proposition 7]{Wallman}. We also mention that this could also be handled in simulation by turning off the \textit{optimization level} when passing the job to QASM Simulator.
\end{enumerate}
The rest of the protocol remains the same. The pseudo-code for Ng-URB is provided in Algorithm~\ref{Algo:2}.

By following the rest of the protocol as in URB, we can then extract the unitarity of native gate-specific noise by 
fitting an exponential curve.

 \RestyleAlgo{boxruled}
 \begin{algorithm}
\caption{Pseudocode for Native Gate URB Protocol for single copy implementation} \label{Algo:2}
Sequence lengths (depths) $\mathbb{M}$, Number of iterations (or number of random sequences) $\text{N}_{m}$ for every depth $m \in \mathbb{M}$, Number of samples ${\text{S}}_{m}$ for each depth and iteration, native gate to be benchmarked $\mathcal{G}$ and ideal identity gate (on all qubits) $\mathbb{I}$. Let $d = 2^{n}$, where $n$ is the number of qubits.
\begin{algorithmic}[1]
\Procedure{Ng-URB, $\mathbb{M}$, $ \lbrace \text{N}_{m} \rbrace$}{}
\State \textbf{for all} depth $m \in \mathbb{M}$
\State \hspace{0.2cm} \textbf{repeat} $\text{N}_{m}$ times
 \State \hspace{0.4cm} Compose the sequence $\mathcal{G}_{m} = \mathcal{G} \circ \mathbb{I}...\mathbb{I} \circ \mathcal{G} \circ \mathbb{I} \circ \mathcal{G};$ where $\mathcal{G} \circ \mathbb{I}$ is applied $m$ times (ignore first $\mathbb{I}$)
\State \hspace{0.6cm} \textbf{for all} non-identity Pauli's \textit{P}, \textit{Q} 
\State \hspace{0.8cm} \textbf{repeat} $\text{S}_{m}$ times
\State \hspace{1cm} Prepare pure states corresponding to 
$\rho^{(P)} = \frac{I+P}{d}$ and $\hat{\rho}^{(P)} = \frac{I-P}{d}$ as mentioned in Sec.~\ref{state_prep};
\State \hspace{1cm} Apply $\mathcal{G}_{m}$ to each state;
\State \hspace{1cm} Measure $E^{(Q)} = Q$ large number of times;
\State \hspace{1cm} Estimate the shifted purity for a sequence,  $$ \quad q_{\text{\textbf{j}}} = \frac{1}{d^2 - 1} \sum\limits_{\text{{\tiny P,Q $\neq$ I}}}\{ \text{\text{Tr}}[E^{(Q)}\mathcal{G}_{\textbf{j}}\rho^{(P)}]-\text{\text{Tr}}[E^{(Q)}\mathcal{G}_{\textbf{j}}\hat{\rho}^{(P)}]\}^{2};$$
\State \hspace{0.8cm} Average over all samples $\Tilde{q}_{n} = \frac{1}{\text{\textbf{S}}_{m}}\sum\limits_{s} q_{s}$;
\State \hspace{0.2cm} Average over all iterations $\hat{q}_{m} = \frac{1}{\text{\textbf{N}}_{m}}\sum\limits_{n} \Tilde{q}_{n}$;
\State Fit exponential curve $\hat{q}_{m} = Bu^{m-1}$, where $u$ is unitarity and $B$ is a constant absorbing SPAM.
\EndProcedure
\end{algorithmic}
\end{algorithm}

\subsection{Implementation details}
\label{Implementation_Ng_URB}
We now discuss our implementation of Ng-URB protocol to characterize the noise of IBM-Q processors. In this section, we discuss the details of our implementation and the results of our experiments are discussed in Sec.~\ref{Simulation}.

The unitarity experiments we perform are simulations using QASM simulator of IBM-Qiskit. We use the NoiseModel.from\_backend(backend) method provided by Qiskit Aer~\cite{IBMQ}, which approximates the noise model of a real backend device for single and two-qubit native gates of the actual processor. We treat this noise as the average noise of a native gate. It simulates the actual noise imparted due to the gate operation on qubits, in real backends, by considering an effective depolarising channel followed by thermal relaxation of the qubit. Qiskit provides the user, the ability to choose whether to simulate actual gate error as a depolarising channel or the thermal relaxation in qubits or both. We ultimately perform experiments by not considering the thermal relaxation so that the approximate noise simulated is a close approximation of noise-induced solely by the gate operation. Hence, we perform  Ng-URB with ideal qubits, and the only source of noise in the qubits, is due to the noisy native gate operation on them. Qiskit also allows the user to apply custom unitary gates by specifying the custom gate's matrix representation.

The native gate set for both Burlington (5-qubit) and Melbourne (15-qubit) backends consists of the set \{CNOT, $I$, $U2$, $U3$\}, where $I$ is the identity gate, and $U2$ and $U3$ correspond to single qubit rotations parameterized by two and three angles respectively. Fig.~\ref{sim_burlington} and~\ref{sim_melbourne} provide the details of the results of the experiments of Ng-URB. We now elaborate on the actual implementation details of our experiment.

For each processor and native gate chosen, we apply a custom \textit{unitary} gate which performs the (ideal) identity operation, followed by that native gate, repeated depth many times in each sample and iteration. We remark here, that this custom unitary gate is \textit{not} decomposed into native gates since Qiskit allows us to define custom \textit{basis\_gates} or gates from which we want our compiled circuit to be executed and we add this unitary gate (essentially, the non-noisy identity gate) and the native gate chosen, to our set of \textit{basis\_gates}. This means, in our compiled circuit, the depth of the circuit is composed of the ideal unitary gate performing identity operation and the noisy native gate chosen which will be the sole source of noise in the circuit. We can be sure of (ideal) identity operation as a custom \textit{unitary} gate because no IBM-Q device (real backend) contains \textit{unitary} gate in its \textit{basis\_gates} set and since Qiskit can only simulate the backend noise of native gates of a given processor, therefore, the custom \textit{unitary} gate we construct is ideal.

We also make sure to include this custom unitary gate in our set of basis gates so that it does not decompose into a noisy native identity (id) gate of the real backend. We choose to apply (ideal) identity gate and native gate to be benchmarked in an interleaved manner because internally Qiskit tries to optimize the circuit when it compiles the circuit to be run. Since we explicitly want the same noisy gate to be applied many times for each depth in an experiment, we specifically want Qiskit to \textit{not} optimize the circuit and remove the number of duplicate gates. This case specifically arises for the cx gate, for which Qiskit does not apply any cx in the compiled version if, the depth is even and applies a single cx if the depth is odd. By interleaving, an ideal identity gate, we disarm this optimization behavior.

\begin{figure*}[t!]
\centering
 \subfloat[]{{\includegraphics[width=0.5\textwidth]{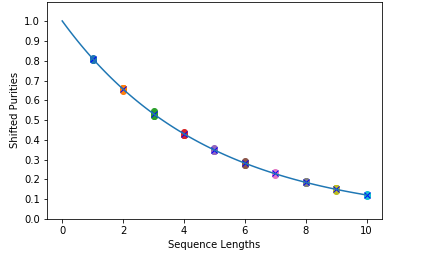} }}
\subfloat[]{{\includegraphics[width=0.5\textwidth]{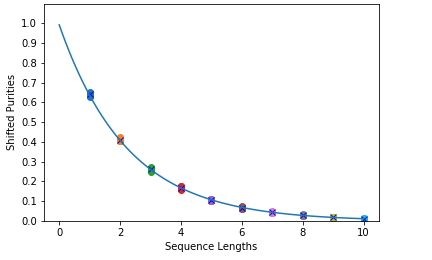} }} \\
\subfloat[]{{\includegraphics[width=0.5\textwidth]{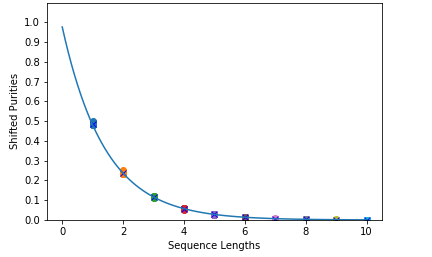} }}
\subfloat[]{{\includegraphics[width=0.5\textwidth]{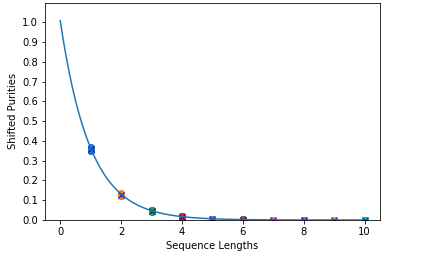} }}
\caption{URB curves showing the relationship between average shifted purities and sequence lengths (in Clifford unitaries) as a result of simulations of URB single copy protocol for completely depolarising noise with different values of parameter $p$, applied to single qubit register. (a) Depolarising parameter $p=0.9$, estimated unitarity $u=0.81015$. (b) Depolarising parameter $p=0.8$, estimated unitarity $u=0.64081$. (c) Depolarising parameter $p=0.7$, estimated unitarity $u=0.49238$. (d) Depolarising parameter $p=0.6$, estimated unitarity $u=0.36072$. For each simulation, the sequence lengths were taken to be $m \in \mathbb{M}$ $=  \left\lbrace 1,2, \dots ,10 \right\rbrace $, number of iterations were 15 and for each iteration, the sample size was 5. Overall variance in the estimated unitarities, in the experiments, are $4.495 \times 10^{-6}$, $6.953 \times 10^{-6}$, $1.272 \times 10^{-5}$ and $1.408 \times 10^{-5}$ respectively.}
\label{sim_dep}
\end{figure*}

\section{Simulation Results of m-URB and Ng-URB protocols in IBM-Q} \label{Simulation}
In this section, we provide the results of our implementations of our two modified versions of URB, namely m-URB and Ng-URB protocol. The results for the simulations involving custom noise models are documented in Sec.~\ref{Noise_sims} whereas, the results for simulations with approximate noise in real IBM-Q devices are provided in Sec.~\ref{NativeURBResults}. We then comment on our results and discuss inferences in Sec.~\ref{discussion}.

\subsection{Results of m-URB for Custom Noise simulation}
\label{Noise_sims}
In this section, we discuss the results of implementing the m-URB protocol discussed in Sec.~\ref{state_prep} and~\ref{measure} for two well-known noise models, namely the (completely) depolarising channel and the bit-flip channel for which, we derived their theoretical unitarity values in Sec.~\ref{derivation}. 

For each noise model, we simulate it using the Qiskits Aer simulator. Note that, these noise models are toy models to simulate the gate error for a gate set. In our case, they simulate the noise associated with a Clifford gate. In our implementation, each Clifford gate is first applied, i.e., we apply the sequence of gates corresponding to the decomposition of the Clifford gate specified by the QASM string generated with the help of the QuaEc library, followed by noisy identity gate modeling the specific noise channel in consideration.

\subsubsection{Depolarising Channel}
Figure~\ref{sim_dep} shows the exponential curves plotted for single-qubit m-URB implementation with noisy simulation of (completely) depolarising noise per single-qubit Clifford gate for different values of $p$. We take the depths or the number of Clifford unitaries applied to be $m \in \mathbb{M}$ $=  \left\lbrace 1,2, \dots ,10 \right\rbrace $. For each depth, we run $15$ iterations and take $5$ samples for given depth and iteration. In each plot, the line represents the best fit curve given by 
\begin{equation}
\Tilde{\text{q}}_{m} = Bu^{m-1},    
\end{equation}
where $u$ is the estimated unitarity from the experiment and $m$ is the number of noisy gates applied and the dots are the shifted purities for different iterations and given depth. The concentration of the dots along the line indicates that the variance in the shifted purities is quite low and is centered around the average shifted purity. Fig.~\ref{sim_dep} (a), (b), (c), and (d) show the fit curves for depolarising parameters $p$ as $0.9, 0.8, 0.7,$and $0.6$ respectively. The unitarity for each case was found to be $0.81015$, $0.64081$, $0.49238$, and $0.36072$ respectively which are close to the theoretical values for each case. Overall variance in the estimated unitarities, in the experiments mentioned above, are $4.495 \times 10^{-6}$, $6.953 \times 10^{-6}$, $1.272 \times 10^{-5}$ and $1.408 \times 10^{-5}$ respectively. We remark that the empirical unitarity values were found to be close to the theoretical values with a precision error in the order of $\sim 10^{-3}$.

\subsubsection{Bit-flip channel}
\begin{figure*}[t!]
\centering
\subfloat[]{{\includegraphics[width=0.5\textwidth]{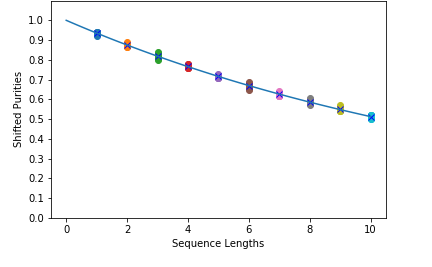} }}
\subfloat[]{{\includegraphics[width=0.5\textwidth]{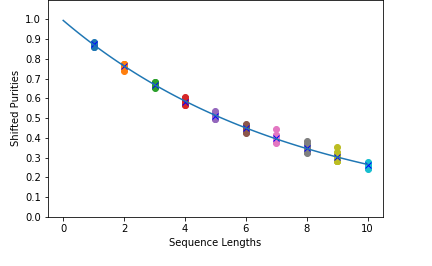} }} \\
\subfloat[]{{\includegraphics[width=0.5\textwidth]{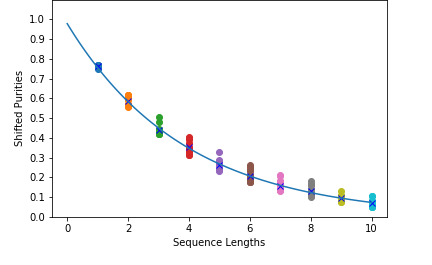} }}
\subfloat[]{{\includegraphics[width=0.5\textwidth]{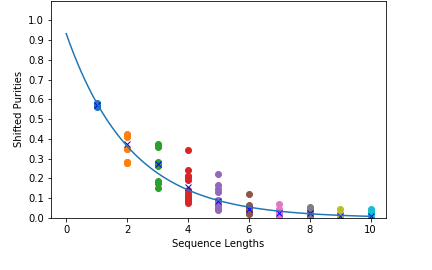} }}
\caption{URB curves showing the relationship between average shifted purities and sequence lengths (in Clifford unitaries) as a result of simulations of URB single copy protocol for the bit-flip channel with different values for probability $p$, applied to single qubit register. (a) For $p=0.975$, the estimated unitarity was found to be $u=0.935424$. (b) For $p=0.95$, the estimated unitarity was found to be $u=0.876434$. (c)  For $p=0.9$, the estimated unitarity was found to be $u=0.772098$. (d) For $p=0.8$, the estimated unitarity was found to be $u=0.624513$. For each simulation, the sequence lengths were taken from $\mathbb{M}$, the number of iterations was 15 and for each iteration, the sample size was 5. Overall variance in the estimated unitarities, in the experiments, are $9 \times 10^{-7}$, $1.775 \times 10^{-5}$, $7.21 \times 10^{-5}$ and $1.81 \times 10^{-3}$ respectively.}
\label{sim_bit_flip}
\end{figure*}

The results of our experiments are shown in Fig.~\ref{sim_bit_flip}. Similar to the depolarising channel, we take the depths or the number of Clifford unitaries applied to be $m \in \mathbb{M}$. For each depth, we run $15$ iterations and take $5$ samples for a given depth and iteration. Fig.~\ref{sim_bit_flip} (a), (b), (c), and (d) show the fit curves of a bit-flip channel, with different values of probability $p$ that the qubit state does not `flip' as $0.975, 0.95, 0.9,$ and $0.8$ respectively. The unitarity was found to be $0.935424$, $0.876434$, $0.772098$, and $0.624513$ respectively which are close to the theoretical values for each case. Overall variance in the estimated unitarities, in the experiments mentioned above, are $9 \times 10^{-7}$, $1.775 \times 10^{-5}$, $7.21 \times 10^{-5}$ and $1.81 \times 10^{-3}$ respectively.

In the above experiments, we find that in extreme conditions, when the noise strength is considerably high, it leads to an overestimation of unitarity when m-URB protocol is used to calculate it. For instance, in the depolarizing channel case, the regime of $p < 0.5$ or in the bit-flip channel, the probability of survival, i.e., the probability that the noise channel will not disturb the state is less than $0.9$. This is largely because of the very weak signal strengths, i.e., the shifted purity values for a modest increase in depth size. As it is evident from Fig.~\ref{sim_dep} and~\ref{sim_bit_flip}, the shifted purities for depth sizes of more than $5$, in the strong regimes for respective noise channels, the shifted purities are close to zero. This induces error in the fitting of the exponential curve and as a result, the empirical unitarity is overestimated by the protocol. However, in the high unitarity regimes, i.e., for the cases in which the shifted purities are sufficiently larger than zero, the results of the m-URB protocol are validated by the theoretical prediction with sub-percent variance in the results. We conclude that our experiments are completely consistent with the theoretical prediction of unitarity and hence, establish the soundness of our implementation.

\subsection{Ng-URB implementation in IBM-Q processors}
\label{NativeURBResults}
In Sec.~\ref{Noise_sims}, we have provided evidence for the soundness of our implementation of m-URB by comparing it with theoretical values. In this section, we present the results of experiments, in which we calculate the unitarity of the approximate (simulated) noise map generated naturally in the operation of native gates present in two of the quantum processors of the IBM-Q devices suite, namely Burlington (5-qubit) and Melbourne (15-qubit).

\subsubsection{5-qubit burlington processor}
\begin{figure*}[t!]
\centering
\subfloat[]{\includegraphics[width=0.5\textwidth]{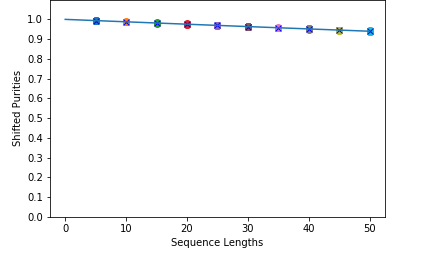} }
\subfloat[]{{\includegraphics[width=0.5\textwidth]{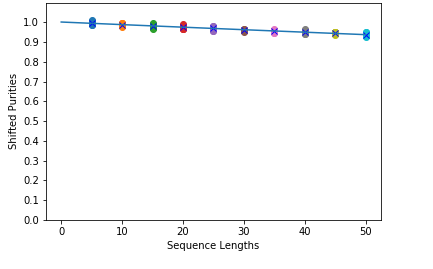} }}\\
\subfloat[]{{\includegraphics[width=0.5\textwidth]{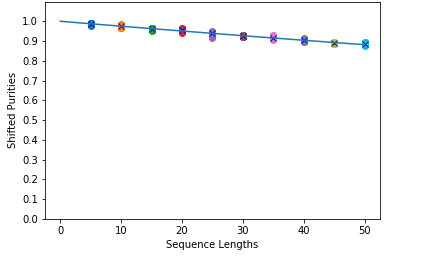} }}
\subfloat[]{{\includegraphics[width=0.5\textwidth]{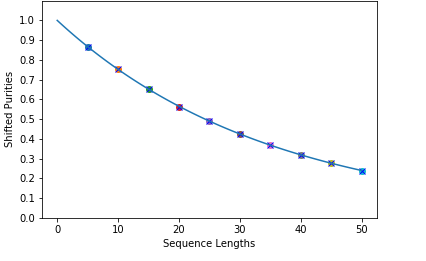} }}
\caption{Results of experimental implementation of a variant of Standard URB procedure, to benchmark unitarity of native gates of Burlington processor (display\_name: ibmq\_burlington). URB curves associated with experiments done with native gates chosen as Identity (id), $U2$ (u2), $U3$ (u3), and CNOT (cx) are shown in Fig (a),(b),(c), and (d) respectively. The depths used for each benchmarking experiment were for identity and CNOT gates, $m \in \mathbb{M}$ $=  \left\lbrace 1,2, \dots ,10 \right\rbrace $ and for $U2$ and $U3$ gates, $m \in \mathcal{M} = \{5k \mid k \in \{1,2,3...10\}\}$. Number of iterations for each benchmarking experiment $= 15$ and number of samples $= 5$. Unitarity of the approximate noise map of id, u2, u3, and cx was found to be 0.998763, 0.998683, 0.997480, and 0.971898 respectively. Overall variance in the estimated unitarities, in the experiments, are $3.229 \times 10^{-9}$, $1.413 \times 10^{-7}$, $4.985 \times 10^{-8}$, and $6.894 \times 10^{-9}$ respectively over 15 iterations of repeating the entire experiment for id, u2 and u3 gates whereas 10 iterations were made for cx gate.}
\label{sim_burlington}
\end{figure*}

\begin{figure*}[t!]
\centering
\subfloat[]{{\includegraphics[width=0.5\textwidth]{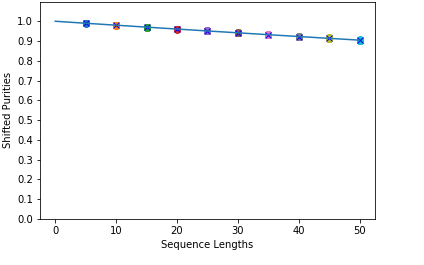} }}
\subfloat[]{{\includegraphics[width=0.5\textwidth]{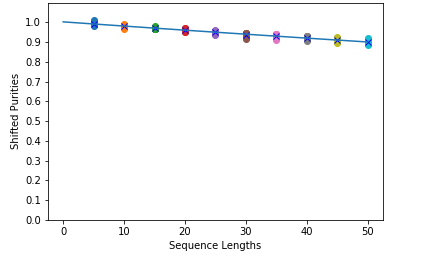} }} \\
\subfloat[]{{\includegraphics[width=0.5\textwidth]{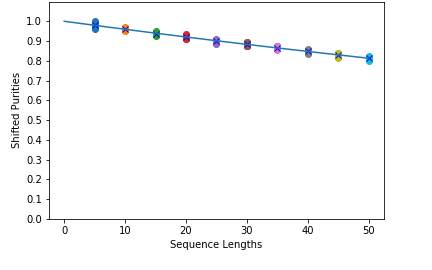} }}
\subfloat[]{{\includegraphics[width=0.5\textwidth]{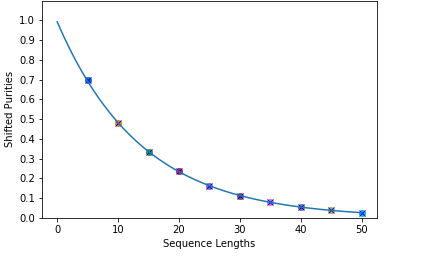} }}
\caption{Results of experimental implementation of Ng-URB procedure, to benchmark unitarity of native gates of Melbourne processor (display\_name: ibmq\_16\_melbourne). URB curves associated with experiments done with native gates chosen as Identity (id), $U2$ (u2), $U3$ (u3), and CNOT (cx) are shown in Fig (a),(b),(c), and (d) respectively. The depths used for each benchmarking experiment are, for identity and CNOT gates  $m \in \mathcal{M} = \{5k \mid k \in \{1,2,3...10\}\}$ and for $U2$ and $U3$ gates, $m \in \{ 1,2,3,4 \}$. Number of iterations for each benchmarking experiment $= 15$ and number of samples $= 5$. Unitarity of the approximate noise map of id, u2, u3, and cx was found to be 0.997995, 0.997861, 0.995872, and 0.930389 respectively. Overall variance in the estimated unitarities in the experiments are $2.778 \times 10^{-9}$, $2.928 \times 10^{-8}$, $2.755 \times 10^{-8}$, and $1.314 \times 10^{-7}$ respectively over 15 iterations of repeating the entire experiment for id, u2 and u3 gates whereas 10 iterations were made for cx gate.}
\label{sim_melbourne}
\end{figure*}

Figure~\ref{sim_burlington} shows the exponential curves obtained for each of the native gates of a 5-qubit Burlington device. For each native gate, we took the sequence lengths or depths to be $m \in \mathcal{M} = \{5k \mid k \in \{1,2,3...10\}\}$. For each depth, the number of iterations performed was 15 and the number of samples was taken to be 5. The optimization level in the \textit{QASM Simulator} is set to 0 (no optimization) and the thermal relaxation of the qubits is ignored for the simulations. Figure~\ref{sim_burlington} (a), (b), (c), and (d) are the fit curves obtained for the identity gate (id), single qubit gate with two parameters (u2), single qubit gate with three parameters (u3) and CNOT (cx) gates respectively of the Burlington device. The estimated unitarities for the native gates were found to be 0.998763, 0.998683, 0.997480, and 0.971898 respectively.  The unitarities for the identity and single-qubit gates are as high as expected, however, the unitarity for the CNOT gate is lower than the single qubit gates though sufficiently high.

\subsubsection{15-qubit melbourne processor}

Figure~\ref{sim_melbourne} shows the exponential curves obtained for each of the native gates of a 15-qubit Melbourne device. For each native gate, we took the sequence lengths or depths to be $m \in \mathcal{M}$ where $\mathcal{M} = \{5n \mid n \in \{1,2,3...10\}\}$. For each depth, the number of iterations performed was 15 and the number of samples was taken to be 5. The optimization level in the \textit{QasmSimulator} is set to 0 (no optimization) and the thermal relaxation of the qubits is ignored for the simulations. Fig~\ref{sim_melbourne} (a), (b), (c), and (d) are the fit curves obtained for the identity gate (id), single qubit gate with two parameters (u2), single qubit gate with three parameters (u3) and CNOT (cx) gates respectively of the Burlington device. The estimated unitarities for the native gates were found to be 0.997995, 0.997861, 0.995872, and 0.930389 respectively. The unitarities for the identity and single-qubit gates are high as expected, however, the unitarity for the CNOT gate is lower than the single qubit gates and is also lower than the unitarity of the CNOT gate for the Burlington device.

In Sec.\ref{discussion}, we discuss the results obtained for the Burlington and Melbourne devices and comment on the magnitude of the estimated unitarity values and its role in detecting cross-talk.

\begin{table}[b]
\caption{\label{IBM_summary}
Table summarising the Native Gate URB experiments on 5-qubit and 15-qubit IBM-Q processors.}
\centering
\begin{ruledtabular}
\begin{tabular}{ c | c | c }
Native Gate & Burlington (5-qubit) & Melbourne (15-qubit)\\
\hline
Identity (`id')&\mbox{0.998763}&\mbox{0.997995}\\
Single-Qubit (`u2')&\mbox{0.998683}&\mbox{0.997861}\\
Single-Qubit (`u3')&\mbox{0.997480}&\mbox{0.995872}\\
Two-Qubit (`cx')&\mbox{0.971898}&\mbox{0.930389}\\
\end{tabular}
\end{ruledtabular}
\end{table}

\subsection{Discussion on Ng-URB Results and detecting cross-talk}
\label{discussion}
In this section, we comment on the results and discuss the extent of the cross talk in the IBM-Q processors which we have considered in this paper. Since the cross-talk errors are correlated~\cite{Cross_talk}, the cross-talk between target qubits (which the gate operation intended to act on) and neighboring qubits (extraneous qubits affected by the gate operation) can induce incoherent errors in the target qubits. Conversely, the target qubits of a given gate operation can be subjected to incoherent noise after the gate is applied, either due to the decoherence of the targets qubits themselves or due to the cross-talk caused by the gate operation itself. We have already ensured that the simulation of the backend noise is executed on the ideal qubits with no thermal relaxation errors. This rules out the decoherence of the target qubits themselves as the source of incoherence in the noise channel.

We conclude from the results summarised in Table~\ref{IBM_summary} that there is a significant cross-talk inducing thr incoherent errors in the case of two-qubit CNOT gate for both 5-qubit and 15-qubit processors. For the rest of the native gates, for both the processors, we conclude that there is little incoherence in the noise and practically the noise remains unitary for low depths. However, for large circuit depths of more than $30$ the incoherence in the noise associated with single qubit gates for both processors becomes significant, considering they only act on a single qubit.

We also conclude that the cross talk induced by the CNOT gate for the 15-qubit processor induces more incoherence in the noise than the 5-qubit one by observing that the unitarity of the CNOT gate in 15-qubit is about $0.04$ less than that of the 5-qubit processor. We remark here that we are only able to detect cross-talk and its effect on the target qubits in terms of incoherence induced in the noise channel through our Ng-URB protocol. We cannot comment on the further characterization of the cross-talk based on the obtained results. Although, quantifying and characterizing the cross-talk based on the concept of unitarity and estimating a figure of merit by employing a randomized benchmarking type protocol might be an interesting question to consider. Nevertheless, the least we can reason is, that simple connectivity and less number of qubits induce less incoherence as a result of cross-talk, while rich connectivity and/or more number of qubits (which is what we see in the case of the 15-qubit device) induces more when applying the same entangling gate such as the CNOT gate. This is however true, only when the quality of qubits in both the processors are comparable, i.e., either they are the ideal qubits or their decoherence times are of the same order.

\subsection{Comparison between Sheldon \textit{et al}'s~\cite{Sheldon} work and our work}
\label{Comparison with Sheldon}
In this section, we compare and contrast our approach of detecting cross talk (or rather, in general, non-unitary errors) with that proposed by Sheldon \textit{et al}.~\cite{Sheldon}. We also discuss the limitations of our proposed protocol from a more practical standpoint.
\begin{enumerate}
    \item \textit{Firstly}, we perform our experiments in a simulator with simulated average noise for custom noise channels and for native gate noise. Dirske \textit{et al}.~\cite{Wallman} provide bounds on the number of iterations and samples to be taken while performing URB in practice, to estimate unitarity that is SPAM independent. We claim that the same is true for our protocol, and similar rigorous bounds for the same are the subject of future research and we only focused on the simulation of the protocol. In fact, we suspect that, if our proposal is implemented as it is, the protocol will underestimate the unitarity because of the additional noise from the noisy gates which are required to prepare the initial states and rotation of basis for measurement as discussed in Sec.~\ref{state_prep} and Sec.~\ref{measure} apart from the target noise to be benchmarked. The remedy is that, we have to perform many more experiments to even out the effects of SPAM.

\item \textit{Secondly}, we use unitarity as our metric to characterize the average noise associated with a gate set (as in m-URB) or a single native gate (as in Ng-URB). We emphasize that unitarity is definitive in the sense that, there exists a rigorous protocol that requires exponential curve fitting to estimate it and the more unitarity is less than one, the more non-unitary is the noise. However, the approach of~\cite{Sheldon} depends on the best-fit curve obtained by plotting the fidelity vs. the number of times the target gate is composed with itself. An immediate fact we notice is that URB and its variants that we propose are only capable of detecting non-unitary error and measuring the extent to which it is non-unitary, whereas, through iterative randomized benchmarking, detection of both unitary and non-unitary errors can be achieved. 

\item \textit{Thirdly}, like many other SRB protocols, URB also assumes that the gate noise to be benchmarked is gate independent, also known as Markovian Noise~\cite{Preskill}. Though for other SRB protocols, modifications taking into account the gate dependence of the noise (non-Markovian) have been proposed. To the best of our knowledge, there do not exist any such modification for URB. In practice, gate-dependent noise for example, can arise from amplitude fluctuations in the local oscillator, an amplifier, or other microwave electronics along the control line of a superconducting qubits-type processor. In general, if the noise is gate-dependent, then the noise due to a target gate is different when the given gate is applied after a sequence of noisy gates as compared to its standalone application. For example, even if the noise in the target gate is unitary in a standalone setting and the same is true for all other gates in the sequence applied in an experiment, if the unitary errors for all the gates in the sequence don't add up coherently, then the combined noise map for the sequence is given by a mixed unitary channel, which is in general non-unitary in nature. Therefore, if the gate dependence is taken into account, then through URB, one will find that the average noise of a gate set is incoherent (non-unitary) but in reality all the gates have only unitary errors which do not add up coherently. Hence, this assumption of gate independence limits the proper characterization of the noise which we aim to perform when applied in practice. 
\end{enumerate}

\section{Conclusion} \label{Conclusion}
We propose a modified version of  URB protocol~\cite{Dirkse, Wallman}, namely, m-URB protocol to experimentally verify the significance of the URB protocol. In these experiments, our control over the exact compilation and actual implementation on the IBM-Q devices is to the extent of simply performing Native Gate URB or Ng-URB successfully. We propose more detailed experiments as future work, in which the user chooses the qubits present on the device (based on the decoherence times of the qubits) as the working qubits and performs URB on these qubits, so that, full characterization of noise can be performed on a given processor, considering its restricted connectivity. Both standard URB and Ng-URB, require the whole Pauli group to be accessed every time we try to benchmark the coherence in the noise, which scales exponentially with the number of qubits. We propose the modification of Ng-URB so that it can be practically tractable also as a future problem. We hope that this can be of potential use to quickly empirically calculate a figure of merit that can give an idea about the amount of cross-talk and incoherence in the quantum computation performed on a given available device or archetype.

\appendix

\section{Significance of URB as a complementary protocol for SRB} \label{Significance}

In this section, we discuss how the definitions of unitarity given in Eqs.~\eqref{unitarity1}, \eqref{unitarity2}, and \eqref{unitarity3} to capture the deviation of a noise channel from being a unitary channel and thereby measuring the `coherence' of a noise channel. Specifically, we will focus on the form of unitarity defined in Eq.~\eqref{unitarity3} and establish the theoretical correctness of URB protocol, specifically by drawing similarity between Eq.~\eqref{unitarity3} and the definition of shifted purity mentioned in Algorithm~\ref{Algo:1}. Next, we compare URB with SRB and its other variants and conclude by highlighting the role of URB, complementary to SRB and other variants in the theory of benchmarking and error correction.\\

Let, $\mathcal{E}$ be a completely positive (CP) map from the space of quantum states in $\mathcal{H}$ to itself. A density operator $\rho$ can be expanded in the Pauli operator basis $\{\sigma_{i}\}_{i = 0}^{d^{2} - 1}$, where $\sigma_{0} = \mathbb{I}$, as follows: (here, $d$ is the dimension of $\mathcal{H}$ and for now, $d = 2^{n}$, $n$ being the number of qubits)
\begin{equation} \label{rhoexpansion}
    \rho = \sum\limits_{i = 0}\limits^{d^{2}-1} \rho_{i}\sigma_{i}.
\end{equation}
We can then write $\mathcal{E}$ as a $d^{2} \times d^{2}$ matrix. Assuming, that the $\mathcal{E}$ is also trace preserving, the matrix for $\mathcal{E}$ looks like
\begin{gather}
\mathcal{E} = 
    \begin{pmatrix}
    1 & 0 & 0 & \dots & 0\\
    \vdots & & \ddots & \\
    \mathcal{E}_{n} & & & \mathcal{E}_{u} & &\\
    \vdots & & & & \ddots \\
    \end{pmatrix}.
\end{gather}
Here, $\mathcal{E}_{n}$ (of dimension $d^{2}-1 \times 1$) is the non-unital part of $\mathcal{E}$, since, if this vector is non-zero then the map $\mathcal{E}$ is said to be non-unital, i.e., doesn't map identity ($\sigma_{0}$) to itself. On the other hand, $\mathcal{E}_{u}$ (of dimension $d^{2}-1 \times d^{2}-1$) is the unital part of $\mathcal{E}$. It has been shown that, $\mathcal{E}_{u}$ is   orthogonal matrix (for, real quantum channels i.e., $\mathcal{E}$ maps Hermitian operators to Hermitian operators) if and only if $\mathcal{E}$ is a unitary channel or mathematically, $\mathcal{E}$ is of the form $U\rho U^{\dagger}$ for some unitary operator $U$ over the Hilbert space $\mathcal{H}$~\cite{Preskill}. Wallman \textit{et al.} in~\cite[Proposition 1.]{Wallman}, prove that the definition of unitarity defined in Eq.~\eqref{unitarity1} is equivalent to:
\begin{equation} \label{unitarity4}
    u(\mathcal{E}) = \frac{\text{\text{Tr}}[\mathcal{E}_{u}^{\dagger}\mathcal{E}_{u}]}{d^{2}-1}.
\end{equation}
Here, we note that, if the noise channel $\mathcal{E}$ is unitary then, since $\mathcal{E}_{u}$ has to be orthogonal, we see that $u(\mathcal{E}) = 1$. The converse of this statement is also true and in fact, for channels other than unitary channels, Wallman \textit{et al.} in~\cite[Proposition 7.]{Wallman}, show that $u(\mathcal{E}) < 1$. We conclude that, in fact, unitarity is a viable figure of merit, to measure the deviation in the behavior of a noise channel from being a unitary channel i.e., it is an indicator of `coherence' in the noise channel.

We now discuss, how the URB protocol mentioned in Algorithm~\ref{Algo:1} lets us empirically calculate, the unitarity for arbitrary noise channels. By a slight algebraic manipulation, the equivalence of the form of Eq.~\eqref{unitarity3} and Eq.~\eqref{unitarity4} can be established as follows. Firstly, note that, if a density operator $\rho$ is expressed as mentioned in Eq.~\eqref{rhoexpansion}, we can write the action of $\mathcal{E}$ on $\rho$ in an algebraic manner as
\begin{equation} \label{noise_expansion}
    \mathcal{E}(\rho) = \sum\limits_{i,j}(\mathcal{E}_{ij}\rho_{j})\sigma_{i}.
\end{equation}
Now, Eq.~\eqref{unitarity3} can be written as
\begin{equation}
\begin{aligned}
u(\mathcal{E}) =& \frac{1}{d^{2} - 1}\sum\limits_{\sigma, \tau \in \mathbb{P}^{*}}\Big\{\text{\text{Tr}}[\tau\mathcal{E}(\sigma)]\Big\}^{2}\\ 
=& \frac{1}{d^2-1}\sum\limits_{i,j=1}\limits^{d^2-1}\Big\{\text{\text{Tr}}[\sigma_{j}\mathcal{E}(\sigma_{i})]\Big\}^{2}\\
=& \frac{1}{d^2-1}\sum\limits_{i,j=1}\limits^{d^2-1}\Big\{\text{\text{Tr}}\Big[\sigma_{j}\sum\limits_{k=0}\limits^{d^2-2}\mathcal{E}_{u_{ki}}\sigma_{k}\Big]\Big\}^{2} \hspace{0.1cm} [\text{from Eq.~\eqref{noise_expansion}}] \\
=& \frac{1}{d^2-1}\sum\limits_{i,j=1}\limits^{d^2-1}\Big\{\sum\limits_{k=0}\limits^{d^2-2}\mathcal{E}_{u_{ki}}\text{\text{Tr}}[\sigma_{j}\sigma_{k}]\Big\}^{2}.
\end{aligned}
\end{equation}
Since, $\text{\text{Tr}}[\sigma_{j}\sigma_{k}] = \delta_{jk}$, (here, $\delta_{ab} = 1$ if and only if $a = b$ and is zero otherwise and using the fact that, $\sigma_{i} \in \mathbb{P^{*}} \text{ (normalized Pauli group) } \hspace{0.05cm} \forall \hspace{0.05cm} i$)
\begin{equation}
u(\mathcal{E}) = \frac{1}{d^2-1}\sum\limits_{i,j=0}\limits^{d^2-2}(\mathcal{E}_{u_{ji}})^{2}
= \frac{1}{d^2-1}\text{\text{Tr}}\left[\mathcal{E}_{u}^{\dagger}\mathcal{E}_{u}\right].
\end{equation}
Here, the last equality is true for real noise channels.

Now, we show that the expression for shifted purity specified in Algorithm~\ref{Algo:1} is equal to the unitarity of a composed noise map formed by applying the sequence of gates to the qubits, upto a constant factor. By slightly modifying the expression for shifted purity given by
 \medmuskip=1mu
\thinmuskip=1mu
\thickmuskip=1mu 
\begin{equation}
\begin{aligned}
q_{\text{\textbf{j}}} =&  \frac{1}{d^2 - 1} \sum\limits_{\text{P,Q $\neq$ I}}\left\lbrace  \text{\text{Tr}}\left[E^{(Q)}\mathcal{G}_{\textbf{j}}\rho^{(P)}\right]-\text{\text{Tr}}\left[E^{(Q)}\mathcal{G}_{\textbf{j}}\hat{\rho}^{(P)}\right]\right\rbrace^{2}\\
=& \frac{1}{d^2 - 1} \sum\limits_{\text{P,Q $\neq$ I}}\left\lbrace  \text{\text{Tr}}\left[Q\mathcal{G}_{\textbf{j}}\left(\frac{\mathbb{I}+P}{d}\right)\right]-\text{\text{Tr}}\left[Q\mathcal{G}_{\textbf{j}}\left(\frac{\mathbb{I}-P}{d}\right)\right]\right\rbrace ^{2}\\
=& \frac{4}{d^2 - 1} \sum\limits_{\text{P,Q $\neq$ I}}\left\lbrace  \text{\text{Tr}}\left[Q\mathcal{G}_{\textbf{j}}(\frac{P}{d})\right]\right\rbrace^{2}.
\end{aligned}
\end{equation}
 \medmuskip=2mu
\thinmuskip=2mu
\thickmuskip=2mu 
More precisely, since, $P$ and $Q$ here are ordinary Pauli gates, by normalizing them (with respect to Hilbert-Schmidt norm), after renaming variables, the final expression for shifted purity looks like
\begin{equation}
\begin{aligned}
q_{\text{\textbf{j}}} =& \frac{4}{d^2-1} \sum\limits_{\sigma,\tau \in \mathbb{P}^{*}}\Big\{ \text{\text{Tr}}\left[\tau\mathcal{G}_{\textbf{j}}(\sigma)\right]\Big\}^{2}.
\end{aligned}
\end{equation}
Here, we see that, shifted purity for a sequence $\mathcal{G}_\textbf{j}$ is equal to the unitarity of the noise map resulting from applying the sequence $\mathcal{G}_\textbf{j}$ (assuming, the gates in $\mathcal{G}_\textbf{j}$ are noisy) upto a constant factor. We note that this constant factor is consumed in the constant $B$ when extracting $u$ by fitting the exponential curve. By extracting the unitarity from the exponential curve mentioned in Algorithm~\ref{Algo:1}, we get an estimate for average unitarity. This is the same as the shifted purities averaged over all sequences up to a constant factor. For each depth value the average shifted purity is the average unitarity of the noise channel, composed with itself depth many times. By using any number of depth values, we essentially try to minimize the variance in estimation. We conclude the significance and correctness of URB protocol as it is proposed in~\cite{Wallman} with the modification in state preparation and measurement for \textit{single copy implementation} suggested in~\cite{Dirkse}.
\section{Comparison between URB protocol with SRB protocol and its variants}
Now, we try to compare and contrast the URB protocol with the SRB protocol and its other variants. 
\begin{enumerate}
    \item \textit{Firstly}, in SRB protocol, after the sequence of gates (preferably from a unitary 2-design) are applied, an inverting gate is applied, so that ideally, the quantum operation is equivalent to applying an identity channel. We are interested in the average survival probability of a given input state, under the action of the averaged noise map of the gate set. Hence, we require efficient implementation (and reversal) of unitary gates in the gate set, since, errors in compiling inefficient circuit implementations might affect the average infidelity. However, we see that, in URB protocol, there is no such need for an inverting gate and the gates can be applied arbitrarily, without keeping track of the gates applied for the inversion step, unlike what we see in SRB.\

\item \textit{Secondly}, in SRB, we start with the vacuum state i.e., the state, $\ket{0}^{\otimes n}$ for $n$ qubits and experimentally determine the probability that the input state survives the noisy channel, by measuring the counts for $\ket{0}^{\otimes n}$ averaged over many iterations. Therefore in SRB, we have to essentially prepare a single state and observe the occurrences of a single state of the computational basis, in the output distribution. On the other hand, we notice that in URB (both for single copy and two copy implementations) there exists an optimal input state and measurement operator (which is a Pauli operator) so that the shifted purity values are of maximum strength (i.e., they are sufficiently larger than 0)~\cite{Dirkse}. While we can always rotate our basis to account for the expectation value of any Pauli observable at the time of measurement, the optimal input states prescribed by the algorithm are in general mixed states for both types of implementation and hence, pose a challenge for experimental implementation of the protocol.\\

\item Now, we try to relate the two figures of merit used in SRB and URB. Wallman \textit{et al.} in~\cite[Proposition 8]{Wallman}, deduce the following upper bound on average gate fidelity $F_{\text{avg}}$, in terms of unitarity ($u$) of the average noise map of the gate set
\begin{equation}
    \Big(\frac{dF_{\text{avg}}-1}{d-1}\Big)^{2} \leq u.
\end{equation}
This inequality shows that, if $u$ is strictly less than 1 (the noise is non-unitary), then the maximum (non-trivial) average gate fidelity for the gate set can be $\frac{\sqrt{u}(d-1) + 1}{d}$, but if the noise is unitary, i.e., $u = 1$ then, there is no non-trivial upper bound for the average gate fidelity and indeed, with unitary noise, the gate fidelity may be close to zero. A low value for average gate fidelity with respect to a non-unitary noise map, maybe the best case benchmark under given, operating characteristics i.e., noisy gate implementations in the quantum processor under consideration, when the inequality is near saturation. On the other hand, if the noise is unitary, then the same low value of average gate fidelity might highlight only the average quality of computation on the given processor. Therefore, we see that gate fidelity alone is not sufficient to characterize and quantify the performance of a noisy computation and we require the characterization of the noise in the gate operations to reach a conclusion. Hence, URB protocol must be treated as a complementary procedure, along with SRB, in order to conclusively benchmark the quality of quantum computation.\\

\item We now discuss, the significance of unitarity in characterizing the diamond distance of a noise channel. In quantum error correction literature, specifically with respect to fault-tolerant computation, the diamond distance of the noise map with respect to the identity channel is regarded as the proper metric for noise strength. Mathematically, the diamond norm of a superoperator $\mathcal{E}$ is defined as~\cite{Preskill}
\begin{equation}
    ||\mathcal{E}||_{\text{\tiny$\Diamond$}} = \max\limits_{\rho}||\mathcal{E}\otimes\mathcal{I}(\rho)||_{1},
\end{equation}
where $\mathcal{E}$ acts on $\mathcal{H}$ (the system) and $\mathcal{I}$ is the identity map acting on $\mathcal{H^{'}}$ (the environment) and maximization is with respect to density operators on the combined Hilbert space $\mathcal{H}\otimes\mathcal{H^{'}}$.
The diamond distance of a noise channel $\mathcal{N}$ from the identity channel $\mathcal{I}$ is then given by~\cite{Preskill}
\begin{equation}
    D_{\text{\tiny$\Diamond$}}(\mathcal{N}) := ||\mathcal{N} - \mathcal{I}||_{\text{\tiny$\Diamond$}}.
\end{equation}
We note here, that the diamond distance quantifies the worst-case error rate of a noise channel with respect to the identity channel and the average gate infidelity quantifies the average error rate of the noise channel. Experimentally estimating the diamond distance is hard, but it can be bounded using experimental benchmarking protocols. Kueng \textit{et al.}~\cite{Kueng}, gave the following bounds on the diamond distance $D_{\text{\tiny$\Diamond$}}$ of a noise channel $\mathcal{N}$, whose average gate infidelity, $r(\mathcal{\mathcal{N}}) = \frac{(d-1)(1 - p)}{d}$, where $p$ is the benchmarking parameter (fit parameter) in SRB and unitarity of $\mathcal{N}$ is $u$
\begin{equation}
    \frac{\sqrt{(d^2-1)(1-2p+u)}}{2d} \leq D_{\text{\tiny$\Diamond$}} \leq \frac{d\sqrt{(d^2-1)(1-2p+u)}}{2}.
\end{equation}
We would like to conclude this section with an important corollary of the above inequality, that if the noise channel is unitary (i.e., $u = 1$), then, we see that the worst case error rate of the noise channel, $D_{\text{\tiny$\Diamond$}}$ scales as $\sqrt{r(\mathcal{N})}$, whereas if the channel is far from unitary (i.e., $u < 1$), for instance, a completely depolarizing channel, with depolarizing parameter $p$, $D_{\text{\tiny$\Diamond$}}$ scales as $r(\mathcal{N})$. Therefore, unitarity and average fidelity together can provide insights into the worst-case error rate of a noise channel which is the theoretical metric used for benchmarking fault-tolerant computation.
\end{enumerate}

\end{document}